\documentclass[fleqn,usenatbib]{mnras}

\usepackage{newtxtext,newtxmath,ulem}
\usepackage{caption}
\usepackage{multirow}
\usepackage{amsmath}
\usepackage{cases}
\usepackage{subcaption}

\usepackage[T1]{fontenc}
\newcommand{\angstrom}{\mbox{\normalfont\AA}}
\DeclareRobustCommand{\VAN}[3]{#2}
\let\VANthebibliography\thebibliography
\def\thebibliography{\DeclareRobustCommand{\VAN}[3]{##3}\VANthebibliography}

\usepackage{graphicx}	
\usepackage{amsmath}	
\usepackage{tabularx}
\usepackage{dcolumn}
    \newcolumntype{d}[1]{D{.}{.}{#1}}

\usepackage{xcolor}
\usepackage{color, soul}

\usepackage{hyperref}
\usepackage{subfiles}

\newcommand{\sqcm}{cm$^{-2}$}  
\newcommand{\kms}{$\rm km~s^{-1}$} 
\newcommand{\lya}{Ly$\alpha$}

\newcommand{\HI}{\mbox{H\,{\sc i}}}

\newcommand{\OI}{\mbox{O\,{\sc i}}}
\newcommand{\OII}{\mbox{O\,{\sc ii}}}

\newcommand{\OVI}{\mbox{O\,{\sc vi}}}

\newcommand{\CIV}{\mbox{C\,{\sc iv}}}

\newcommand{\SiII}{\mbox{Si\,{\sc ii}}}

\newcommand{\NI}{\mbox{N\,{\sc i}}}

\newcommand{\NV}{\mbox{N\,{\sc v}}}

\newcommand{\rewf}{$W_{r,500}$}

\newcommand{\logm}{${\rm log}_{10}(M_*/\rm M_{\odot})$}
\newcommand{\Msun}{$\rm M_{\odot}$}

\title[Neutral hydrogen distribution around low-$z$ galaxies]{MUSEQuBES: Mapping the distribution of neutral hydrogen around low-redshift galaxies} 

\author[S. Dutta et al.]{Sayak Dutta,$^{1}$\thanks{E-mail: sayak18@iucaa.in} 
Sowgat Muzahid,$^{1}$
Joop Schaye,$^{2}$
Sapna Mishra,$^{1}$ 
Hsiao-Wen Chen,$^{3}$ 
Sean Johnson,$^{4}$ 
\newauthor 
Lutz Wisotzki,$^{5}$ 
and 
Sebastiano Cantalupo$^{6}$ 
\\ ~\\ 
$^{1}$Inter-University Centre for Astronomy \& Astrophysics, Post Bag 04, Pune, India, 411007 \\ 
$^{2}$ Leiden Observatory, Niels Bohrweg 02, 2333 CA Leiden, Netherlands \\  
$^{3}$ Department of Astronomy \& Astrophysics, The University of Chicago, Chicago, IL 60637, USA \\ 
$^{4}$ Department of Astronomy, University of Michigan, 1085 S. University Ave, Ann Arbor, MI 48109, USA \\ 
$^{5}$Leibniz-Institute for Astrophysics Potsdam (AIP), An der Sternwarte 16, D-14482 Potsdam, Germany \\ 
$^{6}$Department of Physics, University of Milan Bicocca, Piazza della Scienza 3, 20126, Milano, Italy \\ 
}

\date{Accepted XXX. Received YYY; in original form ZZZ}

\pubyear{2023}

\begin{document}
\label{firstpage}
\pagerange{\pageref{firstpage}--\pageref{lastpage}}
\maketitle

\begin{abstract}
We present a detailed study of cool, neutral gas traced by \lya\ around 4595 $z<0.5$ galaxies using stacks of background quasar spectra. The galaxies are selected from our MUSEQuBES low-$z$ survey along with data from the literature. These galaxies, with a median stellar mass of \logm\ $= 10.0$, are probed by 184 background quasars giving rise to 5054 quasar-galaxy pairs. The median impact parameter is $b = 1.5$~pMpc (median $b/R_{\rm vir}=10.4$) with 204 (419) quasar-galaxy pairs probing $b/R_{\rm vir} < 1 (2)$. We find excess absorption out to at least $\approx 15 R_{\rm vir}$ transverse distance and $\approx \pm 600$~\kms\ along the line of sight. We show that the median stacked profile for the full sample, dominated by the pairs with $b > R_{\rm vir}$, can be explained by a galaxy-absorber two-point correlation function with $r_0 = 7.6$~pMpc and $\gamma = -1.57$ . There are strong indications that the inner regions ($\lesssim R_{\rm vir}$) of the rest equivalent width profile are better explained by a log-linear (or a Gaussian) relation whereas the outer regions are well described by a power-law, consistent with galaxy-absorber large-scale clustering. 
Using a sub-sample of 339 galaxies (442 quasar-galaxy pairs, median $b/R_{\rm vir} = 1.6$) with star formation rate measurements, we find that the \lya\ absorption is significantly stronger for star-forming galaxies compared to passive galaxies, but only within the virial radius. The \lya\ absorption at $b\approx R_{\rm vir}$ for a redshift-controlled sample peaks at $M_* \sim 10^9~ {\rm M_{\odot}}~ (M_{\rm halo} \sim 10^{11}~ \rm  M_{\odot})$.

\end{abstract}




\section{Introduction}

It is now well established that the luminous parts of galaxies are surrounded by a large reservoir of low-density, diffuse gas called the circumgalactic medium (CGM). The CGM serves as a bridge between galaxies and the intergalactic medium (IGM). However, there is no consensus on the actual extent of this medium surrounding galaxies. 

Absorption line spectroscopy of bright background sources such as quasars paved the way to study these elusive media, which are otherwise difficult to detect in emission. The advancement of sensitive space-based detectors such as the Cosmic Origin Spectrograph (COS) onboard the Hubble Space Telescope ($HST$) has drastically increased our ability to study the diffuse gas around galaxies over the last two decades. Several surveys  have established the connections between the CGM and many key galaxy properties using absorption line spectroscopy of background quasars \citep[e.g.,][]{Tumlinson_2017,Peroux_2020}. Alongside the rapidly accumulating data from  galaxy surveys to probe the CGM, the advent of new physical prescriptions for gas flows with new numerical methods and faster computers continues to pose fundamental questions about galaxy formation and evolution. Theoretical studies indeed suggest that the distribution of gas and metals surrounding galaxies is intricately linked to processes such as gas accretion, galactic winds, mergers, and stripping that dictate galactic evolution \citep[e.g.,][]{Rahmati_2016,Voort_19,Oppenheimer_20,Appleby_22, Mitchell_22}. 
It is thus crucial to map the distributions of diffuse matter surrounding galaxies with a range of galaxy properties (e.g., stellar mass ($M_*$), star formation rate (SFR), specific-SFR (sSFR)) to gain insight into such physical processes.

The semi-empirical relation between the stellar mass to halo mass ratio ($M_*/M_{\rm halo}$) as a function of $M_{\rm halo}$ shows a peak at $M_{\rm halo}\approx 10^{12}$~\Msun, suggesting that these halos are most efficient in converting baryons into stars \citep[e.g.,][]{Conroy_2006,Behroozi_2013,Behroozi_2019}. It is not well understood why halos of a certain mass have higher star formation efficiencies (SFE). Energy and momentum feedback due to supernovae (SNe) and active galactic nuclei (AGN) is thought to suppress the SFE for the low-mass and high-mass halos, respectively \citep[see e.g.,][]{Bower_2006, Somerville_2008, Crain_2015} . Theoretical models suggest two distinct modes of gas accretion depending on the halo mass: (a) ``hot mode'' for galaxies with $M_{\rm halo}>10^{12}$~\Msun\ and (b) ``cold mode'' for galaxies with $M_{\rm halo}<10^{12}$~\Msun\ \citep[see e.g.,][]{keres, Voort_2011}. The halos of masses $\approx10^{12}$~\Msun\ benefits from both the modes. Thus, the higher SFE for these halos may be related to the gas accretion process. It is therefore essential to probe how the cool, neutral gas reservoir surrounding galaxies changes as a function of stellar/halo mass.

The connection between galaxies and the cool, neutral gas surrounding them has long been a subject of research, both observational and theoretical. \citet[]{Chen_2005} first observed a morphology dependent galaxy--\lya\ absorber cross-correlation function. They found that the cross-correlation function between  \lya\ absorbers (with column density $N(\HI)>10^{14}$~\sqcm) and absorption-line dominated galaxies within a projected distance of $1h^{-1}$~Mpc is  significantly lower compared to emission-line dominated galaxies \citep[see also][]{Chen_09}. On the other hand, later CGM surveys such as COS-Halos did not find any significant difference in the \HI\ content (rest-frame equivalent width (REW or $W_r$) and column density ($N$)) around star-forming and passive galaxies on CGM scales \citep[typically within $0.5 R_{\rm vir}$; see e.g.,][]{Thom_2012,Tumlinson_2013}. Although a small fraction of the passive galaxies in their sample did not exhibit any detectable \lya\ absorption, overall the detected \lya\ absorption was similar around the star-forming and passive galaxies.

Earlier, \citet[]{Chen_98, Chen_01} noted that the $W_r$-profile of \lya\ absorption around galaxies depends on the $K$-band and $B$-band luminosities (proxies for the $M_*$ and recent SFR, respectively). Combining the observational data from the COS-Halos and COS-Dwarf \citep[][]{Bordoloi_2014} surveys, \citet[]{Borthakur_2016} found a strong anti-correlation between \lya\ $W_r$ and sSFR. Additionally, they reported similar exponential scale lengths of the \lya\ $W_r$-profile for passive and star-forming galaxies. The larger dispersion in the radial distribution seen for passive galaxies has been interpreted as an indication of patchiness in their CGM. A positive correlation between \lya\ $W_r$ and stellar mass was reported by \citet[]{Bordoloi_2018}. More recently, \citet[]{Wilde_2021} presented the \lya\ covering fraction- and column density-profiles for three different bins of stellar mass. The covering fraction profiles were found to be very different for the high-mass ($M_*>10^{9.9}$~\Msun) and low-mass ($M_*<10^{9.2}$~\Msun) galaxies in their sample. The \lya\ covering fraction is as high as $\approx90$\% within $R_{\rm vir}$ for the high-mass subsample, but gradually declines to only $\approx20$\% at $b/R_{\rm vir} \approx 6$. The covering fraction for the low-mass sub-sample, on the other hand, remains roughly constant at $\approx40$\% all the way out to $\approx 6 R_{\rm vir}$.  

Although the CGM is thought to play a crucial role in driving galaxy evolution, there is no consensus on the extent of the CGM around galaxies. It is generally thought that the CGM extends out to the virial radius of a galaxy \citep[]{Tumlinson_2017}. However, \citet[]{Wilde_2021} recently argued that the CGM traced by neutral hydrogen extends beyond the virial radius \citep[see also][]{Wilde_2023}. In the study by \citet[]{Prochaska_2011}, the detection rate and $W_r$ of \lya\  absorbers are found to be correlated with galaxies out to at least $1$~pMpc, far beyond the virialized halos of the galaxies. They, however, suggested that the weak \lya\ absorbers arising at large impact parameters ($b$) may be unrelated to the gaseous halos around galaxies, and may be  tracing the large-scale environments in which the galaxies are embedded. \citet[][]{Tejos_14} studied the galaxy--\lya\ absorption correlation function and found that it is significantly different from the galaxy--galaxy autocorrelation function, and the difference is primarily driven by `weak' \HI\ absorbers ($N(\HI\ )<10^{14} ~ \rm cm^{-2} $). They concluded that $>50\%$ of weak absorbers are not correlated with galaxies, and hence, galaxies and these absorbers may not trace the same underlying dark matter distribution. \citet[]{Wakker_2015} used the \lya\ absorption to study nearby galaxy filaments using $HST$/COS spectra of 24 background AGN. They observed a trend of increasing \lya\ equivalent width and line width with decreasing filament impact parameter. The \lya\ absorption detection rate is $\approx80$\% within 500~pkpc (proper kpc) of galaxy filaments, but no absorption is seen at $\gtrsim2$~pMpc. A study with a statistically significant number of quasar-galaxy pairs with a wide range of impact parameters, such as the one presented here, is essential to probe the inner and outer regions of galaxy halos simultaneously. This, in turn, allows one to investigate the extent of the so-called CGM.

In order to build a statistically significant sample of background quasar--foreground galaxy pairs, we compiled data from several low-$z$ CGM surveys in the literature along with our own data obtained from the MUSE Quasar-fields Blind Emitters Survey (MUSEQuBES). MUSEQuBES is a dual MUSE program with 16 quasar fields to study the CGM of low-$z$ galaxies and 8 quasar fields to study the gaseous environments of high-$z$ \lya\ emitters \citep[see][]{Muzahid_20,Muzahid_2021}. Integral field spectroscopy (IFS) with MUSE allows us to search for galaxies, particularly the continuum-faint ones, around background quasars more efficiently than with multi-object spectroscopy (MOS) and conventional long-slit spectroscopy. On the flip-side, MUSE has a relatively small ($1'\times1'$) field of view (FoV) compared to a typical MOS spectrograph. Consequently, the probability of finding massive galaxies in the MUSE FoV is small, as they are rare. Therefore, combining IFS and MOS/long-slit observations of quasar fields provides an optimal way to probe the CGM and large-scale structures around galaxies with a wide range of stellar masses simultaneously.

Although MUSE can detect pure line emitters without any detectable continuum, we restricted our analysis to the continuum-detected galaxies in order to be able to estimate their stellar (halo) mass. Our deep observations with MUSE (2--10 hours of exposure time per field)  enable us to obtain a galaxy sample of relatively low mass (median $M_*\sim 10^{8}~\rm M_{\odot}$) compared to the existing CGM surveys in the literature. The MUSEQuBES galaxies along with the archival galaxy samples allow us to explore the uncharted territory of the CGM around low-mass, intermediate  redshift galaxies (see orange points in Fig.~\ref{fig:prop}).

Instead of the usual practice of identifying individual absorption features in quasar spectra and associating them with foreground galaxies, we used the spectral stacking method to map the \HI\ gas, traced by the \lya\ absorption, in the CGM. 
 The stack of \lya\ absorption at the rest-frame of galaxies provides statistical inference on the mean or median \HI\ absorption around the galaxies without any prior knowledge regarding individual absorption systems. Although the presence of saturated absorbers complicates the possible inference of physical quantities such as column density, the strength of the absorption can be easily determined from the $W_r$ \citep[see e.g.,][]{Steidel_2010}. 
Recently, \citet[]{Ho_21} showed that the common approach of galaxy-absorber association based on LOS velocity cuts (e.g., $\pm$300 \kms\ or $\pm$500 \kms) suffers from projection effects, which are more prominent at larger impact parameters. Spectral stacking enables us to be agnostic about the individual galaxy-absorber associations. Further, stacking a large galaxy sample can significantly improve the spectral signal-to-noise ratio (S/N). Stacking, however, erases the kinematic information of individual absorbers, which provides important insights on gas flow processes in galaxies \citep[e.g.,][]{Bouche_2013,Muzahid_2015}.

The connections between neutral gas and galaxies have been studied for high-$z$ galaxies as well \citep[e.g.,][] {Steidel_2010,Muzahid_2021,Lofthouse_2023}. \citet{Rakic_2012} produced 2D \lya\ optical depth maps as a function of impact parameter and LOS velocity separation for $z \approx 2.3$ star-forming galaxies. Such maps show a clear evidence for redshift-space distortion along the LOS direction which cannot be fully attributed to redshift errors \citep[]{Turner_14}. Moreover, comparing with the {\sc eagle} simulation \citep[]{Schaye_15}, \citet[]{Turner_17} found that infalling gas can account for the redshift-space distortion observed in the optical depth maps. Kinematic information in such maps are a useful tool to understand gas flow processes in galaxies \citep[e.g.,][]{Chen_20}. Here we present \lya\ optical depth maps for the first time for low-$z$ galaxies.

This paper is organized as follows. Section \ref{sec:data} summarizes the data sample used in this work. In Section \ref{sec3}, we provide the absorption data analysis. Section \ref{sec:results} presents the results of this work. Section \ref{disc} presents a discussion of the main results of our analysis followed by a summary in Section \ref{sec:summ}. Throughout this paper, we adopt a $\Lambda$CDM cosmology with $\Omega_{\rm M}=0.3$, $\Omega_{\Lambda}=0.7$, and $H_0 = 70~ \rm km~s^{-1}~Mpc^{-1}$.

\begingroup
\begin{table*}
\caption {\label{tab:table1} Overview of the 16 MUSEQuBES fields. From left to right, the columns show the quasar name, right ascension (J2000), declination (J2000), redshift, $V$ band magnitude, exposure time ($t_{\rm exp}$) for the COS/G130M grating, $S/N$ per resolution element at $\lambda=1250$ $\rm \angstrom$, exposure time for the COS/G160M grating, $S/N$ per resolution element at $\lambda=1650$ $\rm \angstrom$, HST programme ID of the COS observations, exposure time for the MUSE observations, and the effective seeing measured in MUSE cubes at $\lambda$=7000 $\rm \angstrom$.
} 
\begin{tabular}{lcccccrrrrrcc} 
\hline 
 QSO & RA & Dec & $z_{\rm QSO}$ & $m_V$ & \multicolumn{2}{c}{G130M}  & \multicolumn{2}{c}{G160M} & PID & \multicolumn{2}{c}{MUSE observations}   \\ \cline{6-7} \cline{8-9} \cline{11-12} 
 &   (J2000) & (J2000) & &  & $t_{\rm exp}$[h] & $S/N$ & $t_{\rm exp}$[h] & $S/N$ & & $t_{\rm exp}$[h] & Seeing[$''$] \\ \hline 

 HE 0435-5304  & 04:36:50.8   & -52:58:49 & 0.425   & 16.4  & 2.3 & 14.5 & 2.5 & 5.8 & 11520 & 2 & 0.97 \\

 HE 0153-4520  & 01:55:13.2   & -45:06:12 & 0.451   & 15.2  & 1.5 & 28.6 & 1.6 & 17.2 & 11541 & 2 & 0.78\\
 
 RXS J02282-4057  & 02:28:15.2   & -40:57:16 & 0.494 & 14.3 & 1.9 & 41.4 & 2.2 & 21.6 & 11541 & 8 & 0.56\\
 
 PKS 0405-12  & 04:07:48.5   & -12:11:36 & 0.574   & 14.9  & 6.7 & 87.1 & 3.1 & 34.7 & 11508, 11541 & 9.75 & 0.72\\
 
HE 0238-1904  & 02:40:32.6   & -18:51:51 & 0.631   & 15.0  & 4.0 & 32.6 & 2.1 & 20.6 & 11541, 12505 & 10 & 0.76\\
 
3C 57  & 02:01:57.1   & -11:32:34 & 0.669  & 16.4 & 3.0 & 29.1 & 2.4 & 13.9 & 12038 & 2 & 0.70\\
 
PKS 0552-640  & 05:52:24.6   & -64:02:11 & 0.680   & 15.0  & 2.6 & 30.8 & 2.3 & 20.1 & 11692 & 2 & 0.77\\
 
PB 6291  & 01:10:16.3   & -02:18:51 & 0.956   & 17.6  & 5.9 & 13.3 & 5.9 & 7.9 & 11585 & 2 & 1.20\\
 
Q 0107-0235  & 01:10:13.2   & -02:19:53 & 0.958  & 17.8 & 7.8 & 14.8 & 12.3 &11.0 & 11585 & 2 & 1.07\\
 
HE 0439-5254  & 04:40:11.9   & -52:48:18 & 1.053   & 16.1  & 2.3 & 19.9 &2.5 &8.8 & 11520 & 2 & 0.70\\ 
 
HE 1003+0149  & 10:05:35.2  & +01:34:44 & 1.078   & 16.9  & 3.1 & 11.0 & 6.2 & 9.6& 12264 & 2 & 0.90\\
 
TEX 0206-048  & 02:09:30.8   & -04:38:27 & 1.128   & 17.2  & 3.9 & $16.6^a$ & 7.8 & 13.9 & 12264 & 8 & 0.70\\
 
Q 1354+048  & 13:57:26.2  & +04:35:41 &1.234   & 17.2  & 3.9 & 16.9 & 7.8 & 6.9 & 12264 & 2 & 0.56\\
 
Q 1435-0134  & 14:37:48.2   & -01:47:11 & 1.310   & 15.8  & 6.2 & 35.3 & 9.5 & 23.8 & 11741 & 5 & 0.54\\
 
PG 1522+101  & 15:24:24.5   & +09:58:30 & 1.324   & 16.2  & 4.6 & 25.0 & 6.4 & 18.1 & 11741 & 2 & 0.59\\
 
PKS 0232-04  & 02:35:07.2   & -04:02:05 & 1.438   & 16.5  & 4.4 & 19.4 & 6.3& 13.1 & 11741 & 2 & 0.83\\ 
 \hline 
\end{tabular}
\begin{tabbing}
Note-- $^{\mathrm{a}}$This is the S/N at $\lambda=1350$$\rm \angstrom$. Due to a Lyman limit system at $z = 0.390$, there is no flux at $\lambda<$ 1280 $\rm \angstrom$. 
\end{tabbing}
\end{table*}
\endgroup

\section{Data} 
\label{sec:data}

\subsection{Galaxy sample from the MUSEQuBES survey}  
\label{sec:galdata}

The low-$z$ part of the MUSEQuBES survey targeted 16 $z\approx0.5-1.5$ UV-bright quasar fields using VLT/MUSE. The MUSE observations were conducted between September 2014 and April 2017 (ESO programmes 094.A-0131, 095.A-0200, 096.A-0222, 097.A-0089 and 099.A-0159; PI: Schaye), with a total exposure time of 62.75~h. The quasars were selected solely based on the availability of high $S/N$ FUV spectra obtained with $HST$/COS. $HST$ and MUSE observation details of these 16 quasar fields are tabulated in Table \ref{tab:table1}. Each MUSE observation block of 1 h was split into 4 $\times$ 900 s exposures, which were rotated by $90^\circ$ and offset by a small $\approx 1-5''$ shift from each other. The data reduction is performed using the standard MUSE data reduction pipeline \citep[v1.2;][]{Weilbacher_20}, adopting the default (recommended) set of parameters. A few additional reduction procedures using the {\sc CubEx} package \citep[]{Cantalupo_2019} were carried out to improve flat-fielding and sky-subtraction using the {\sc CubeFix} and {\sc CubeSharp} routines, respectively. The procedure is detailed in \citet[]{Borisova_2016}, and more recently in \citet{Muzahid_2021}.

The effective seeing per field, corresponding to the full width at half maximum (FWHM) of a 2D Gaussian profile fitted to a point source at $\lambda = 7000~\rm \angstrom $ in the reduced and combined data cube, varies between $0.54''$ and $1.20''$, but is typically $0.7'' - 0.8''$ (see Table~\ref{tab:table1}). With the MUSE field-of-view (FoV) of $1' \times 1'$ centered on the quasar, we are able to observe a region of $480~\rm pkpc \times 480~$pkpc around the QSO at $z \approx 1$ ($110~\rm pkpc \times 110~$pkpc at  $z\approx0.1$). The field is spatially sampled by a grid of $0.2'' \times 0.2''$ pixels. All MUSE observations were carried out using the standard wavelength range of 4750--9350 \AA, sampled by 1.25 \AA\ spectral pixels. The resolving power ranges from $R\approx$ 1800 at $\lambda = 5000$~\AA\ to $R \approx 3500$ at $\lambda = 9000$~\AA, corresponding to a FWHM of 167~\kms\ to 86~\kms, respectively.


The details of galaxy identifications and galaxy property measurements will be presented in a future work. Here, we briefly outline the main steps. First, we run the {\sc Source Extractor} \citep[SExtractor;][]{Bertin_1996} on the MUSE white-light images using a detection threshold of $1\sigma$ per pixel ({\sc DETECT\_TRESH = 1}) and requiring a minimum number of neighbouring pixels above the threshold of 3 ({\sc DETECT\_MINAREA = 3}). The 1D spectra of the continuum-detected objects, extracted from the MUSE cubes using the {SExtractor}-generated segmentation maps, are then inspected by a modified version of {\sc MARZ} \citep[][]{Hinton_2016} to determine their redshifts based on the spectral features. The redshifts are further refined using a modified version of the code {\sc PLATEFIT} \citep[][]{Brinchmann_2004} by fitting Gaussian profiles to the available emission and absorption line features. Note that the wavelengths in the MUSE data cubes are given in air. We applied appropriate corrections while determining the galaxy redshifts.

The H$\alpha$ or [\OII] (when H$\alpha$ is not covered or not detected at $>3\sigma$) line fluxes returned by {\sc PLATEFIT} are used to estimate the SFRs of the galaxies using the relation from \citet{Kennicutt_1998} or from \citet[][for \OII\ luminosities]{Kewley_2004} adjusted for the \citet[]{Chabrier_2003} initial mass function (IMF). 
 The H$\alpha$ emission-line flux is corrected for dust extinction using the flux ratio of H$\alpha$ and H$\beta$ lines. By comparing H$\alpha$/H$\beta$ to its intrinsic value of 2.85, corresponding to Case B recombination at a temperature of $T \sim 10^4$~K and electron densities of $n_{\rm e} \sim 10^2-10^4 {\rm ~cm}^{-3}$ \citep[][]{Osterbrock_2006}, we derive a correction for the H$\alpha$ flux, assuming a \citet[]{Cardelli_1989} reddening curve. For galaxies with H$\alpha$ coverage beyond the MUSE spectra, we use H$\beta$ to calculate the SFR, under the condition that we can correct the line flux for dust extinction using the H$\beta$/H$\gamma$ ratio. We require that both H$\beta$ and H$\gamma$ be detected with $S/N>3$. We then convert the corrected H$\beta$ flux into the H$\alpha$ flux to obtain the SFR, making use of the known intrinsic ratio between the H$\alpha$ and H$\beta$ fluxes. Dust correction is not performed for galaxies with \OII\ based SFR measurements.

The stellar masses of galaxies are estimated using stellar population synthesis (SPS) code {\sc FAST} \citep[]{Kriek_2009}, which fits SPS templates to a set of photometric flux values. Owing to the lack of ancillary photometric data of the quasar fields in our sample, we constructed 11 pseudo-filters with a width of 400~\AA\ each and spanning the wavelength range from 4800--9200~\AA, after masking the prominent emission lines. We calculated the filter flux by convolving the 1D galaxy spectrum– the same one as used for the redshift determination with MARZ– with a boxcar function centered at $\lambda$ = 5000, 5400, . . . \AA. The halo mass and virial radius, defined as the mass and radius of a spherical region within which the mean mass density is 200 times the critical density of the universe, are estimated from stellar mass and redshift measurements  of galaxies using the abundance matching relation from \citet[]{Moster_2013}

In total, we have 475 galaxies detected in the 16 MUSE fields. Note that these are galaxies detected 3000~\kms\ blueward of the corresponding quasar redshifts. The galaxies have redshifts ranging from $0.05-1.4$, and stellar masses and SFRs  ranging from $10^{6.0}$ to $10^{11.8}~\rm M_{\odot}$ and $10^{-3}$ to $10^{2.7}~\rm M_{\odot}~\rm yr^{-1}$ with the median values of $10^{8.9}~\rm M_{\odot}$ and $10^{-0.7}~\rm M_{\odot}~ yr^{-1}$, respectively. The galaxies have impact parameters from the corresponding background quasars in the range 10--320 pkpc with a median value of 150 pkpc (median $b/R_{\rm vir}=1.7$).

\subsection{Galaxy sample from the literature} 

 Besides the MUSEQuBES galaxies, we combined galaxy samples from six different CGM surveys from the literature; namely COS-Halos \citep[]{Tumlinson_2013},  \citet[]{Liang_14}, COS-Dwarf \citep[]{Bordoloi_2014}, COS-Gass \citep[]{Borthakur_2015}, \citet[]{Johnson_15} (hereafter Johnson+15), and \citet[]{Keeney_2018} (hereafter Keeney+18). Brief summaries of these surveys are presented in Appendix~\ref{sec:A1}-\ref{sec:A6}. Before merging the galaxy catalogs, we confirmed that a given galaxy observed in different surveys was not counted multiple times. To eliminate repetition, we ensured that there are no two galaxies within $1''$ spatially and within 500~\kms\ along LOS distance with each other. In cases of multiple occurrences, we count them only once (60 such cases). This provided us with a large sample of $\sim$9000 galaxies. A significant fraction of galaxies in this sample come from Keeney+18. The redshift and stellar mass distributions of the galaxies in the combined sample are shown in Fig.~ \ref{fig:all_mz_dist}.

\begin{figure*}
\centering
    \centering
    \includegraphics[width=1.0\linewidth]{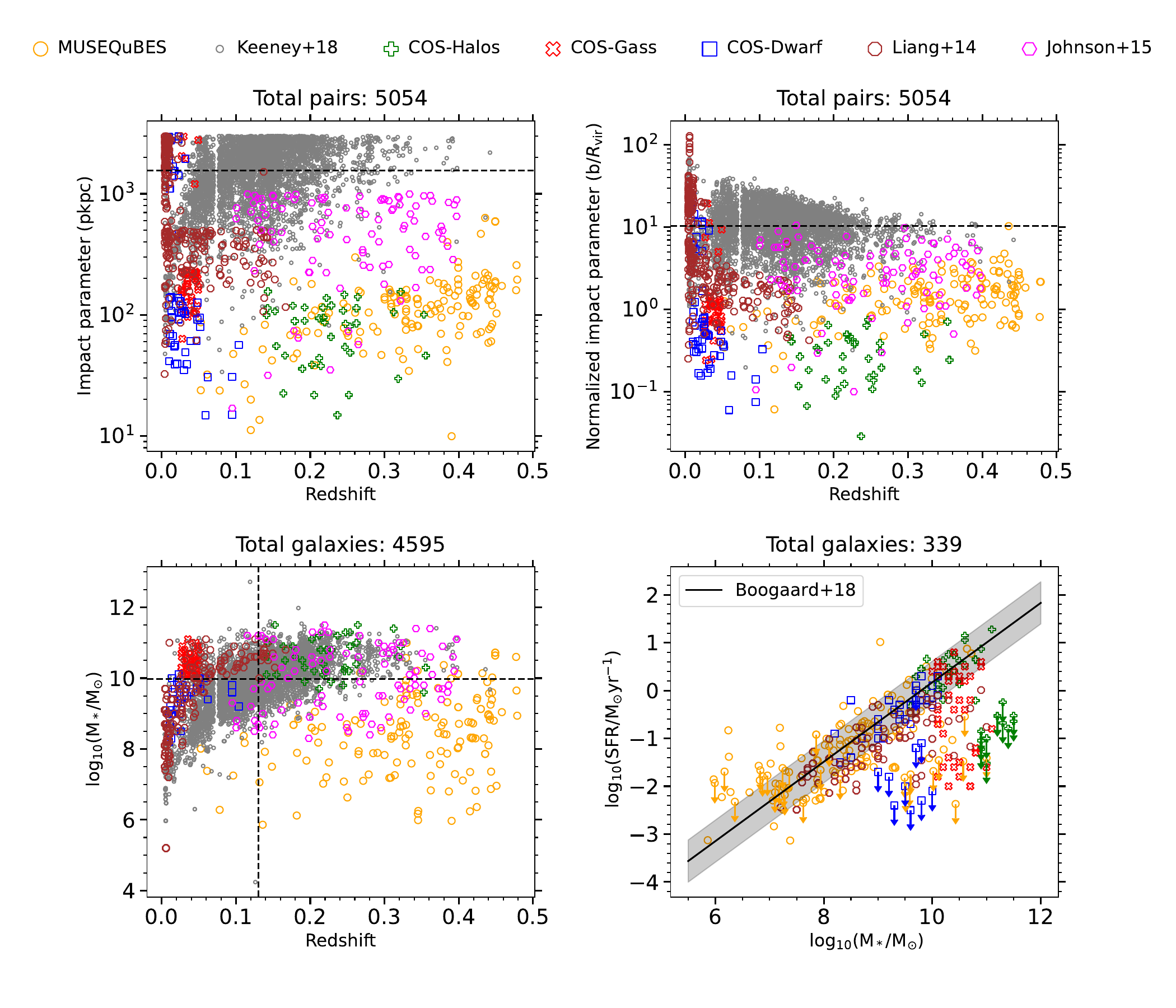}
    \vskip-1.0cm
    \caption{{\tt Top-left} and {\tt top-right:} Impact parameter ($b$) and normalized impact parameter ($b/R_{\rm vir}$) plotted against redshift for the 5054 quasar-galaxy pairs used in this work. The horizontal black dashed lines represent the median $b$ and $b/R_{\rm vir}$ for the complete sample. {\tt Bottom-left:} Stellar mass plotted against redshift of the 4595 galaxies. The black dashed lines indicate the median \logm\ and $z$. {\tt Bottom-right:} SFR plotted against stellar mass for the subsample of 339 galaxies with SFR measurements. The downward arrows indicate upper limits. The black solid line and the shaded region represent the star-forming main sequence relation obtained by \citet[]{Boogaard_18} (at $z=0.1$) and its 1$\sigma$ spread. The data points in all of these panels are color-coded by the survey from which a given galaxy (or a quasar-galaxy pair) is drawn. Noted that \citet{Keeney_2018} and \citet{Johnson_15} did not report SFRs. It is worthwhile to note that the MUSEQuBES data points (orange circles) make up for the uncharted territory of the parameter space covered by the previous CGM surveys (low-mass and higher redshift) and increase the low-mass galaxy sample size significantly.}      
\label{fig:prop}
\end{figure*}

\begin{figure*}
\centering
    \centering
    \includegraphics[width=1.0\linewidth]{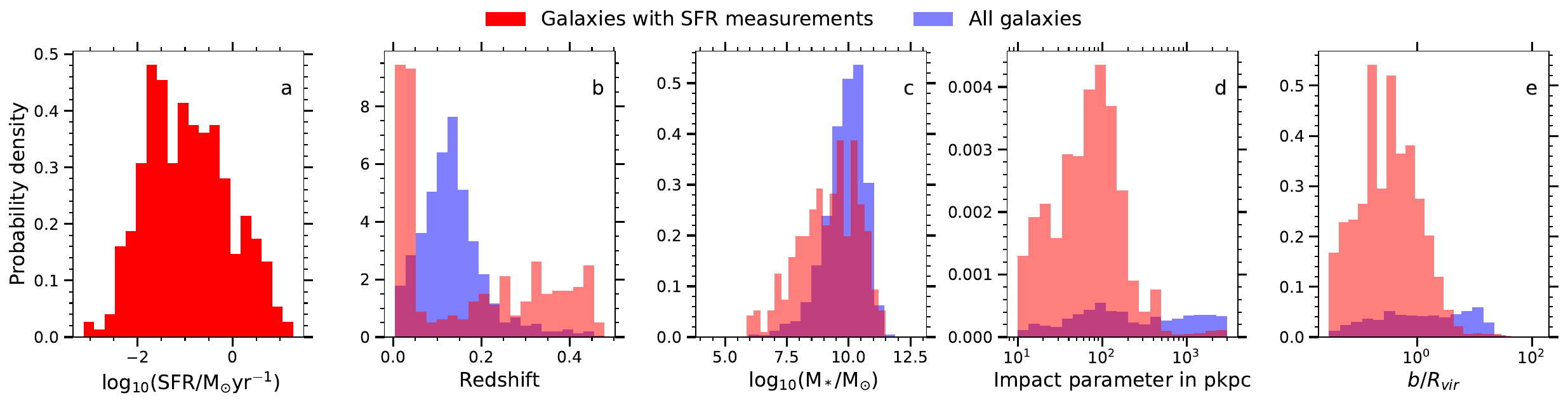}    
    \caption{(a) SFR, (b) redshift, (c) stellar mass, (d) impact parameter, and (e) normalized impact parameter probability density distribution functions for the 4595 galaxies (339 with SFR measurements) in our complete sample. The galaxies with SFR measurements (including upper limits) are shown in red.}    
\label{fig:prop_dist}
\end{figure*}

\subsection{Quasar spectra and continuum fitting}  
\label{sec:qsodata}

Galaxies from the MUSEQuBES survey were detected in 16 fields centered on 16 UV-bright quasars. As mentioned earlier, these quasars were chosen solely based on their FUV brightness and the availability of high $S/N$ COS  spectra. The COS spectra of the quasars that are part of the six studies from the literature are also available in the $HST$ public archive. In total, we obtained 190 COS spectra in reduced form from the $HST$ Spectroscopic Legacy Archive \citep[HSLA;][]{Peeples_2017}. COS has a resolving power of $R\approx 18,000$ (FWHM~$\approx 18$~\kms).

Among the 190 quasars, 135 are observed with both the G130M and G160M gratings covering 1150--1800~\AA. The remaining 55 quasars are observed with the G130M grating only, which has a spectral coverage of 1150--1450~\AA. The G130M and G160M grating spectra of a given quasar are spliced together at the wavelength where the average $S/N$ per pixel becomes equal (typically around 1424 $\angstrom$). The quasar redshifts  range from 0.03--1.88 with a median of $0.43$.

These 190 quasar spectra have a wide range of $S/N$, with some quasars having $S/N$ as low (high) as 1--2 (85--90) per resolution element. The median $S/N$ per resolution element in the G130M for the 190 spectra is $\approx 10$, computed around 1250~\AA. The median $S/N$ for the G160M grating is $\approx 7$, computed near 1650~\AA. Note that we did not impose any $S/N$ cut while selecting the quasars. This is because only 19 (39) spectra have $S/N < 5$ in the G130M (G160M) grating, covering redshifted \lya\ absorption for only 72 (25) galaxies. We verified that excluding these low-$S/N$ spectra does not have a significant effect on our results.

We performed continuum fitting for all of these 190 spectra using a custom-made, semi-automated {\sc python} routine which finds the absorption-free regions in a spectrum in an automated way. First, each spectrum is divided into several chunks and fluxes are sigma clipped with an upper bound of $50\sigma$ (to take into account the emission lines as well). The lower bounds of sigma clipping are varied between 2.5, 2.1, 1.6, and 1.3~$\sigma$ for regions with median $S/N$ (per pixel) of $>10$, $3-10$, $2-3$, and $<2$, respectively. These numbers were chosen after many trials on randomly selected spectra from our sample and by visually inspecting the quality of the continuum model. Next, we increased the chunk size (or decreased the number of chunks) in iterations and repeated the clippings, until the number of clipped pixels converges, or the number of chunks is reduced to 2, whichever happens first. Then, we used spline fitting for the absorption-free pixels to estimate the continuum. The number of `knots' around the known emission lines of \lya\, \CIV\, \NV\, and \OVI\ are increased (if present in the spectra) to take care of larger curvatures.

\subsection{Quasar-galaxy pairs}  
\label{sec:qgp}

Using the 190 quasars in the background of $\approx9000$ galaxies we constructed quasar-galaxy pairs using the following two conditions: (1) The projected distance of a galaxy from the corresponding quasar sightline, $b < 3000$~pkpc. (2) The line of sight (LOS) separation of a galaxy from the background quasar is $>2000$~\kms. The large impact parameter cut-off of 3000 pkpc was chosen to sample the large-scale environments of the galaxies. The limiting LOS separation of 2000~\kms\ from the quasar was chosen to avoid the quasar's proximity zones and associated absorbers arising from the quasars \citep[e.g.,][]{Muzahid_2013}. Using these criteria, we obtained a total of 6020 quasar-galaxy pairs with 5310 unique galaxies. Some of the galaxies in this work are probed by multiple quasars. For example, the COS-GASS and COS-Dwarf surveys are restricted to impact parameters of 250 pkpc and 150 pkpc, respectively. However, a fraction of these galaxies is also probed by quasars at a transverse distance of $\approx1$~pMpc. These provide us with more skewers at larger impact parameters, and hence are included in our studies (see blue squares and red crosses at $\approx1$~pMpc in top-left panel of Fig.~\ref{fig:prop}).
We obtained the redshift, stellar mass, virial radius, and SFR (when available) for all the galaxies from the literature, except for the MUSEQuBES galaxies. There are only 634 galaxies, constituting 882 quasar-galaxy pairs, for which SFRs are known (including upper limits).

Note that the redshifted \lya\ absorption cannot be observed with COS for all of these galaxies as the spectral coverage of COS allows a maximum observable redshift of $0.48$ ($0.19$) for \lya\ for the G160M (G130M) grating. Below, we describe our scheme to obtain the galaxies contributing to the \lya\ stacks.

First, we masked the following regions of the COS spectra:  
\begin{itemize}
    \item 1212--1220~\AA\ -- to exclude the geocoronal Ly$\alpha$  emission and Galactic \lya\ absorption. 
    \item 1301--1307~\AA\ -- to exclude the geocoronal [\OI]   emission.
    \item All wavelengths below the Lyman limit (rest-frame 912~\AA) when a Lyman limit system (LLS; $\log_{10} N(\HI)/\rm cm^{-2} > 17.2$) is present.  
    \item $\pm0.5$~\AA\ around known strong Galactic absorption lines (e.g., \NI, \SiII). At this point, we note that while $\pm0.5$~\AA\ does not exclude all high-velocity clouds (HVCs), we verified that a broader mask of $\pm1.0$~\AA\ does not change any of our conclusions.  
\end{itemize}

The only galaxies for which the redshifted \lya\ wavelengths ($\pm$0.05~\AA) fall within the spectral coverage of the corresponding quasar spectra are considered here. The above conditions led to a total of 5054 galaxy-quasar pairs, with 184 background quasars probing 4595 foreground galaxies. The subsample of galaxies with measured SFR (including upper limits) is reduced to 339 probed by 157 background quasars, giving rise to a total of 442 quasar-galaxy pairs. We note here that a significant fraction (152/442) of them are arising from our MUSEQuBES survey. The SFR is measured (not a limit) for 289 galaxies, constituting 389 galaxy-quasar pairs.

\subsection{Properties of the galaxies contributing to {\texorpdfstring{\lya}{}} stacks}    
\label{subsec:galprop}

In Fig.~\ref{fig:prop} we plot the different galaxy properties against one another. The top-left and top-right panels show the impact parameter and normalized impact parameter ($b/R_{\rm vir}$) of different quasar-galaxy pairs plotted against the redshift of the galaxies. The bottom-left panel shows the stellar mass of the galaxies plotted against redshift. The bottom right panel shows the SFR of the sub-sample of galaxies plotted against stellar mass. The different markers indicate the studies from which the galaxies are drawn. The median (68\% range) $b$ and $b/R_{\rm vir}$ for the 5054 quasar-galaxy pairs used in this study are 1.5~pMpc (0.5--2.5 pMpc) and 10.4 (3.8--17), respectively. There are 204 (419) quasar-galaxy pairs having $b/R_{\rm vir} < 1~(<2)$. The median (68\% range) \logm\ and $z$ for the 4595 unique galaxies are 10 (9.1--10.6) and 0.13 (0.07--0.19), respectively.

The presence of correlations between different galaxy properties is clearly visible in Fig.~\ref{fig:prop}, particularly between the stellar mass and redshift. The Spearman rank correlation test results are summarized in Table~\ref{tab:table_corr} for the full sample and the MUSEQuBES galaxies separately. The redshift and stellar mass are more tightly correlated for the full sample ($r_s=0.48$) as compared to the MUSEQuBES galaxies ($r_s= 0.27$). This is due to the fact that most of the galaxy surveys in the literature are magnitude limited.

For the MUSEQuBES survey, the FoV of MUSE sets the upper limit on the impact parameter at a given redshift, which in turn leads to a tight correlation between redshift and impact parameter for the  MUSEQuBES galaxies ($r_s = 0.52$). However, the combination of different surveys eliminates this strong correlation for the full sample ($r_s =0.24$). The mild anti-correlation seen between normalized impact parameter and redshift is a direct consequence of choosing a fixed upper bound on the impact parameter ($b_{\rm max}= 3$~pMpc).\footnote{Since the galaxies at higher redshifts tend to have higher stellar masses (and hence higher halo masses and larger virial radii), $b/R_{\rm vir}$ values decrease with redshift, leading to the observed anti-correlation.} At this point, we emphasize that the galaxies observed in MUSEQuBES survey (orange circles) populate a region of parameter space ( most prominent in $z-M_*$ plot of Fig.~\ref{fig:prop}) that previous CGM surveys did not cover.

The distributions of individual galaxy properties (SFR, redshift, stellar mass, impact parameter, and normalized impact parameter from left to right) are shown in Fig.~\ref{fig:prop_dist}. 
The distributions corresponding to the complete sample are shown in blue, while the subsample with SFR measurements is shown in red. Among the surveys from the literature used in this work, Keeney+18 and Johnson+15 did not report SFRs for their galaxies. The SFR of the subsample with the rest of the 339 galaxies varies from $10^{-1.82}$ -- $1~\rm M_{\odot}~yr^{-1}$ (68\%) with a median of $0.1~\rm M_{\odot}~yr^{-1}$. {\footnote{The upper limits on the SFR are regarded as measured values for the median and 68\% measurements.}} The range and median redshift for the  galaxy subsample with SFR measurements are consistent with the full galaxy sample. The median stellar mass of the galaxy subsample with SFR measurements ($\log_{10}(M_*/\rm M_{\odot})=9.3$) is lower than for the full sample ($\log_{10}(M_*/\rm M_{\odot})=10.0$). The typical impact parameter of this subsample is also lower than for the full sample (median $b \approx$ 160 pkpc or $b/R_{\rm vir}\approx1.6$ compared to $\approx$~1.5 pMpc or $b/R_{\rm vir}\approx10.4$ for the full sample).

\begingroup
\begin{table}
\centering 
\caption {\label{tab:table_corr}Correlation between galaxy properties} 
\begin{tabular}{cccc}
\hline
 Prop-1 & Prop-2 &  \multicolumn{2}{c}{$r_s^a$} \\ \cline{3-4}  
        &        &  Full Sample & MUSEQuBES     \\ \hline
 $z$  & $b$ & 0.24 & 0.52   \\
 $z$  & $b/R_{\rm vir}$ & -0.13 & 0.31   \\
 $z$  & $M_*$ & 0.48 & 0.27   \\ 
 $b$  & $M_*$ & 0.25 & 0.29   \\ 
 \hline
\end{tabular}
~\\ 
Note-- $^{a}$ Spearman rank correlation coefficient. The $p$-values are $<10^{-10}$ for all cases. 
\end{table}
\endgroup

\section{Absorption data analysis}
\label{sec3}

We study the distribution of neutral hydrogen around the galaxies in our sample by analyzing the \lya\ absorption signal.  Instead of fitting individual \lya\  absorption lines associated with each galaxy, a statistical approach is adopted. We stack the quasar spectra after shifting them to the rest-frames of the corresponding foreground galaxies. Here, we focus on the median-stacked \lya\ absorption.  However, we verified that mean-stacked \lya\ absorption produces consistent conclusions. Spectral stacking is an efficient technique to statistically study the CGM without having to go through rigorous absorption line analysis such as identification, deblending, and profile fitting of individual lines. This method has been successfully used to analyze CGM absorption signals in previous studies \citep[see e.g.,][]{Steidel_2010, Rakic_2012, Turner_14,Chen_20,Muzahid_2021}.

\subsection{Analysis with rest-frame equivalent width}
\label{stack method}

 To obtain the stacked spectra, each normalized quasar spectrum is shifted to the rest-frames of each of the foreground galaxies with $b<3$ pMpc. The median stacked spectrum is then generated by calculating the median flux in line-of-sight velocity bins of 40~\kms. We have verified that our results are insensitive to the bin size. We have not applied any weighting for the  stack, all the galaxy-quasar pairs are treated equally because they provide different and independent probes of the CGM. 
 
  To briefly summarize our analysis procedure: a) we first fit a global continuum to each quasar spectrum and normalize the spectrum by this continuum (see Section~\ref{sec:qsodata}). b) For every foreground galaxy, we select a region of $\pm1500$~\kms\ around the galaxy's redshifted \lya\ wavelength of the normalized spectrum of the corresponding background quasar. c) We obtain the median normalized flux in 40~\kms\ wide velocity bins for a set of quasar-galaxy pairs. This is referred to as the observed median stacked flux profile.  
   d) Next, a local pseudo-continuum is estimated and subtracted out from the observed median stacked spectra to account for the suppression of the overall continuum below unity due to uncorrelated absorbers. The median flux within a LOS velocity window of $\pm2000$ \kms\ for a stack of random redshifts is used to determine the pseudo-continuum (see Section \ref{sec:results} for details). The offset of the observed pseudo-continuum from unity is added to the median stacked spectra to obtain the continuum-subtracted median stacked spectra.
  
  Finally, e) the \lya\ $W_r$s are measured from direct integration of this continuum subtracted median stacked spectra using LOS velocity windows of $\pm300$~\kms\ ($W_{r,300}$) and $\pm 500$~\kms\ ($W_{r,500}$) from the line center. These velocity windows are commonly used in the literature. The errors on $W_r$ measurements are obtained from 1000 bootstrap realizations of the galaxy sample \footnote{   Each bootstrap realization produces a stack of $m$ quasar-galaxy pairs from a sample of $m$ quasar-galaxy pairs, but with replacement.}. We confirmed that convergence is reached in bootstrap distribution for $\gtrsim200$ realizations.   

\begin{figure*}
\centering
\hbox{
\includegraphics[width=0.58\linewidth]{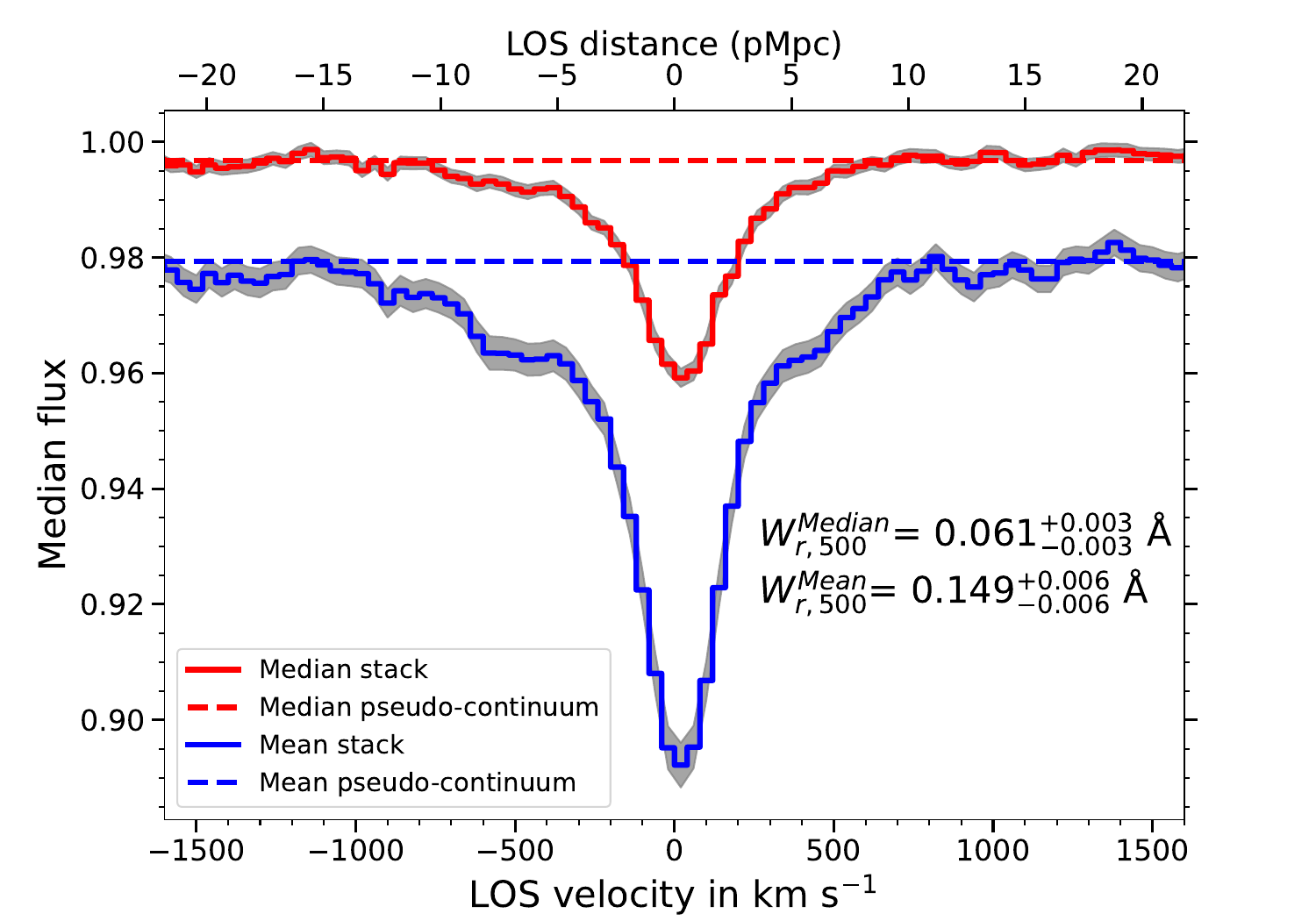}

\includegraphics[width=0.45\linewidth]{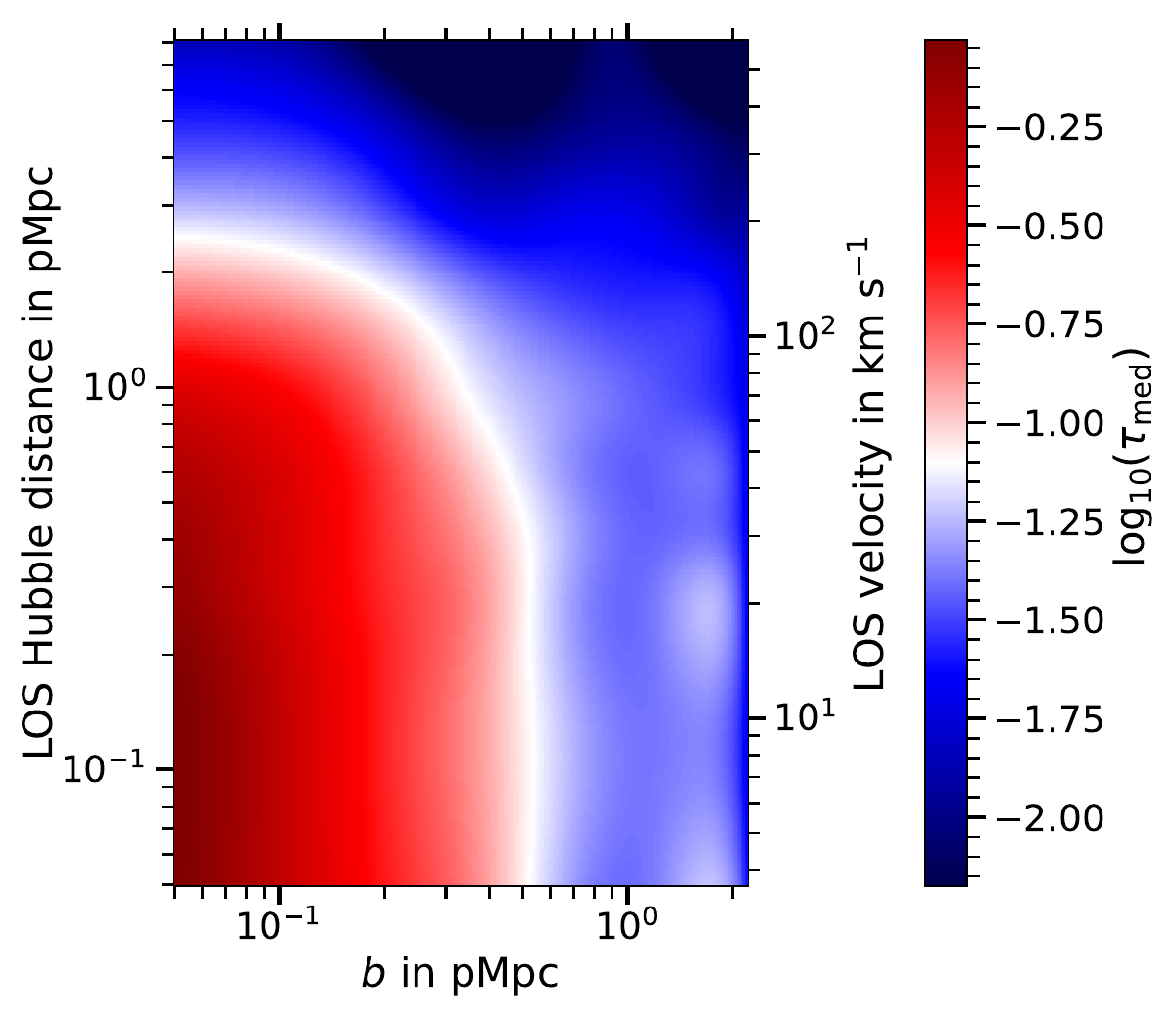}
}  
\caption{ {\tt Left:} The median (red) and mean (blue) stacked \lya\ absorption profile for the full sample in bins of 40~\kms. The mean and median pseudo-continua, shown by the blue and red dashed lines respectively, are obtained from stacks of random redshifts. The shaded regions indicate 68\% confidence intervals of the mean and median flux distributions obtained from 1000 bootstrap realizations of the complete sample. {\tt Right:} Median \lya\ optical depth around galaxies as a function of impact parameter and (absolute) LOS Hubble distance (left y-axis) and LOS velocity (right y-axis). The first bin is confined within 0.1 pMpc, onwards the bin size is 0.17 dex and 0.29 dex for LOS distance and impact parameter, respectively. The map has been smoothed with a Gaussian kernel with standard deviation of half of the bin size. The minimum of the optical depth color scale is set to the value obtained for the stack around random redshifts. The elongation of the signal along the LOS direction indicates the presence of redshift space distortions.}  
\label{fig:all stack}
\end{figure*} 

\subsection{Analysis with pixel optical depth}
\label{od_recovery}

Owing to the large dynamic range compared to the normalized flux, the pixel optical depth ($\tau$) is another useful quantity that has been used to study the connections between galaxies and gas around them \citep[e.g.,][]{Rakic_2012, Turner_14}. We obtained pixel optical depth from the continuum normalized flux $F$ as 
 \begin{equation}
     \tau= -\textrm{ln}(F)
 \end{equation}
 We set a flag value of $\tau=10^{-6}$ for $F>1$. For heavily saturated pixels with $F\leq0$ or $F<\Delta F$ where $\Delta F$ is the error in flux, we set the flag value to $\tau=10^{4}$. The flag values are chosen arbitrarily small and large to ensure that they do not affect the measured median. Similar to our analysis with flux stacking, we masked the aforementioned spectral regions before obtaining the optical depths.

 To produce 2D optical depth maps, we stack the \lya\ optical depth of the absorption associated with galaxies in bins of impact parameter (transverse distance from galaxies) the same way we do as flux stacking. Instead of integrating the median stacked flux profile over the LOS velocity to obtain rest-frame equivalent width, the median optical depth is color-coded as a function of LOS Hubble distance along the y-axis (by converting the LOS velocity to a distance assuming pure Hubble flow at the median redshift of the sample) and transverse distance along the x-axis. 
 These optical depth plots retain information about \lya\ absorption along the LOS direction along with the transverse direction, thus providing insight into the average kinematics of \lya\ absorption of a galaxy sample alongside the absorption strength.

 \begin{figure*}
\centering
    \begin{subfigure}{0.50\textwidth}
        \centering
    
        \includegraphics[width=1.0\linewidth]{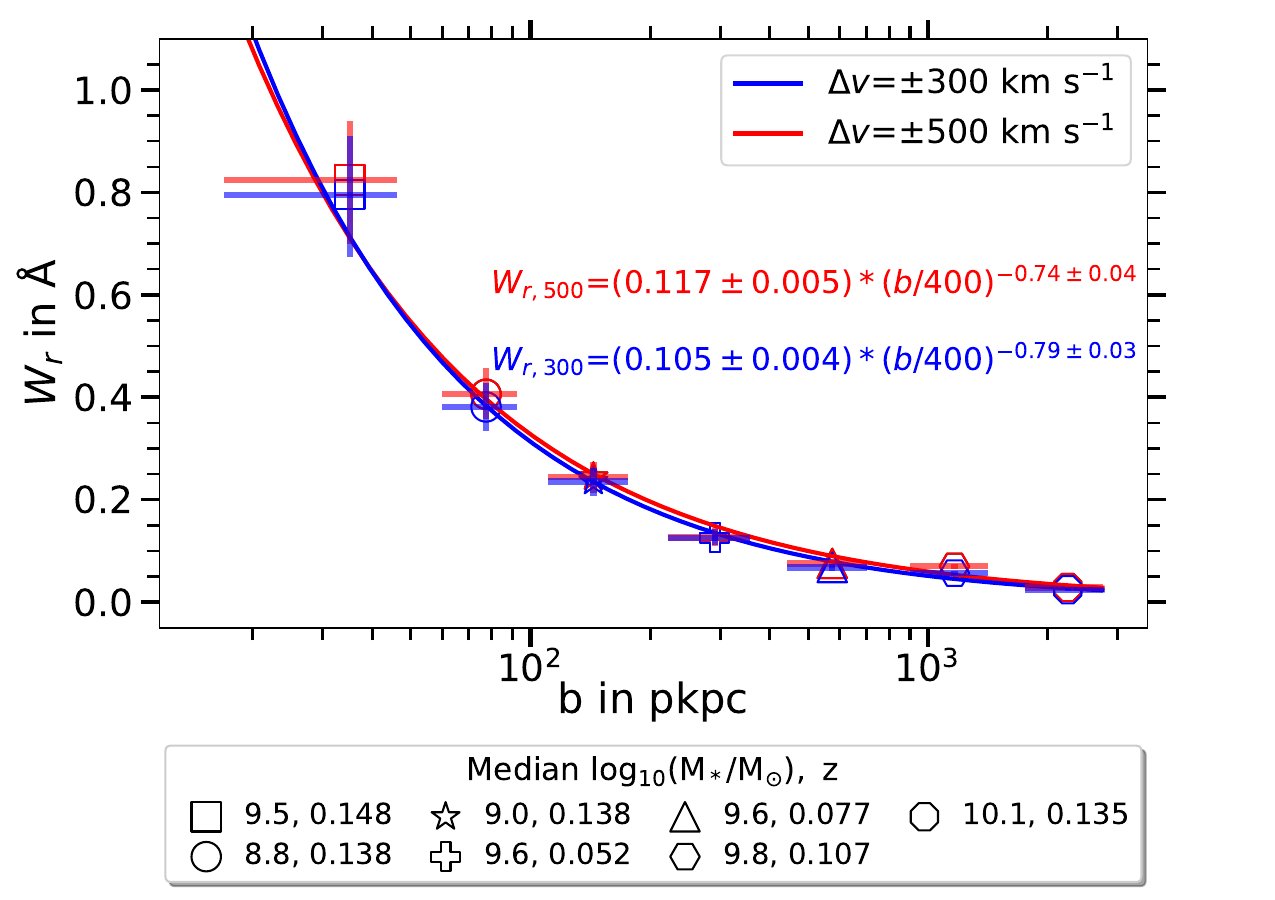}        
       
    \end{subfigure}%
    \begin{subfigure}{0.50\textwidth}
        \centering
      
        \includegraphics[width=1.0\linewidth]{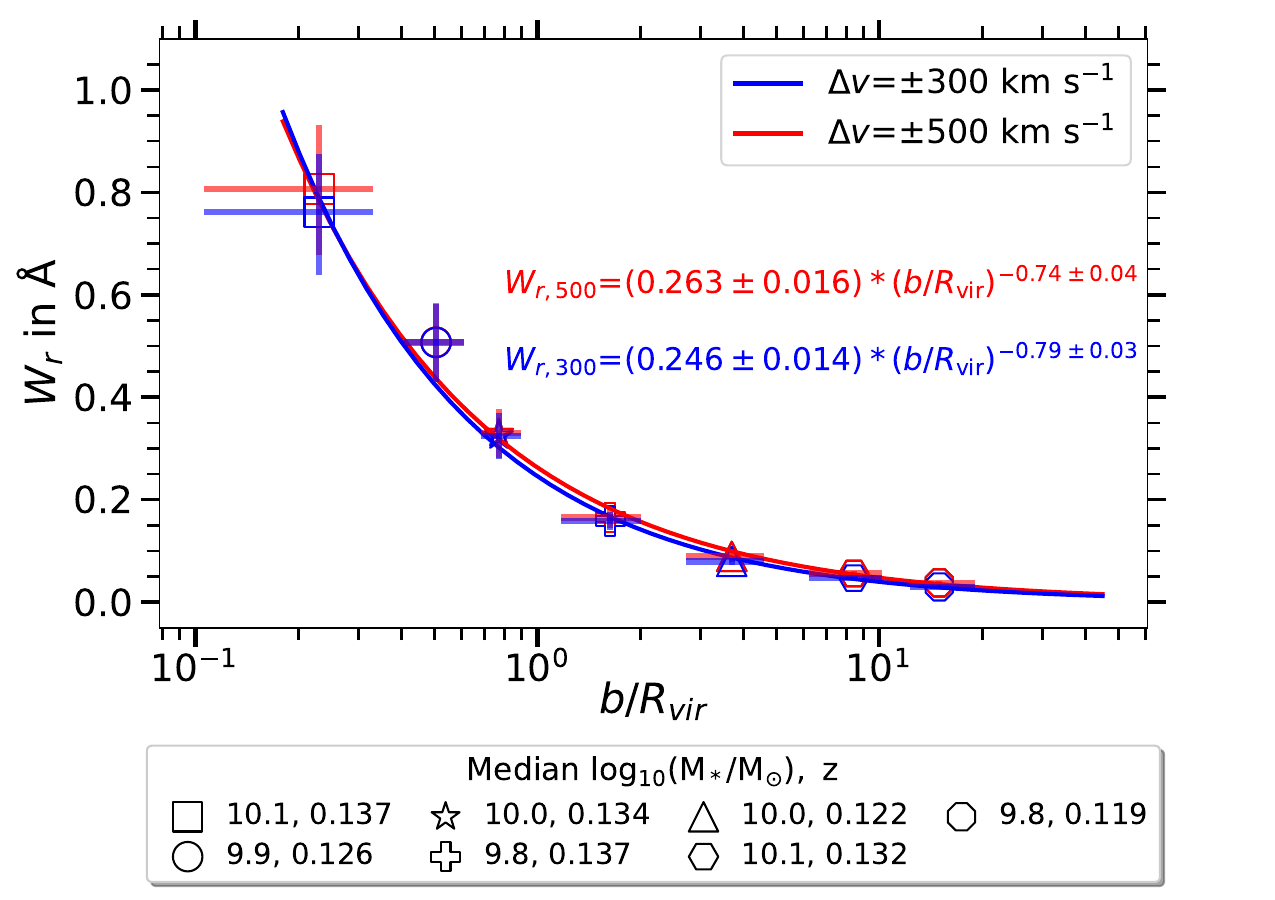}        
        
    \end{subfigure}
\caption{Median Ly$\alpha$ rest equivalent width as a function of physical impact parameter (left) and normalized impact parameter (right). The red and blue points denote rest equivalent widths calculated within velocity windows of $\pm 500$~\kms\ and $\pm300$~\kms, respectively, and are plotted against the median  $b$ ($b/R_{\rm vir}$) of each bin. The error bar on the equivalent width is the $68$\% confidence interval obtained from 1000 bootstrap samples. The error on $b$ ($b/R_{\rm vir}$)  represents the $68$\% percentile of the $b$ ($b/R_{\rm vir}$) distribution in each bin. The median stellar masses and redshifts in each bin are listed in the legends with different markers at the bottom of the plots. The solid blue and red lines in both these plots are the best-fitting power-laws to the blue and red data points, respectively. These best-fit single component power-law relations are provided in the plots.}
\label{fig:b stack}
\end{figure*}

\section{Results} 
\label{sec:results} 

The red (blue) absorption profile in the left panel of Fig.~\ref{fig:all stack} shows the median (mean) stacked  normalized flux for \lya\ as a function of LOS velocity for all galaxies in our sample. The pseudo-continua of the mean and median stacked spectra are lower than the actual continuum level, which is unity for individual normalized spectra. 
Random absorbers with uniform velocity distribution with respect to a galaxy redshift cannot produce any coherent absorption line. They will only suppress the overall continuum below unity, leading to a pseudo-continuum.

In order to model the pseudo-continua we determined the median and mean \lya\ flux around random redshifts. The random redshifts are chosen from a uniform distribution between the lowest and highest galaxy redshifts in each quasar field. The number of random redshifts is kept the same as the number of galaxies in each field.\footnote{As an alternate approach, we obtained random redshifts $\pm 10000-20000$~\kms\ away from each galaxy redshift in each field to ensure that the distributions of galaxy redshifts and random redshifts are the same statistically. There is a negligible difference between the pseudo-continua obtained for these two different cases ($<0.03\%$). We stick to the former approach to obtain pseudo-continua for our entire analysis.} The average of median (mean) flux within LOS velocity of $\pm$2000 \kms\ is used as the median (mean) pseudo-continuum. The 68\% confidence intervals of the median/mean flux, shown by the grey-shaded regions, are calculated from 1000 bootstrap realizations of the complete sample. 

 The observed flux profiles are conventionally normalized by the pseudo-continua before measuring the $W_r$s \citep[see e.g.,][]{Steidel_2010, Prochaska_2013}. 
Instead of normalizing by the pseudo-continuum, we integrate the observed median stacked flux spectrum over the velocity window of $\pm 300$ \kms\ ($\pm 500$ \kms) and subtract out the contribution stemming from the pseudo-continuum within the same velocity window to obtain the $W_r$s. This ensures that the $W_r$ values do not depend on the pseudo-continuum level, which is determined by stochastic absorption. We found that due to the small values of flux decrement, owing to the lower density of the low-$z$ \lya\ forest, the conventional choice of continuum normalization and our adopted choice of continuum subtraction produce consistent results.

The observed $W_{r,300}$ and $W_{r,500}$ for the median stack {\bf for the full sample} are  $0.051_{-0.003}^{+0.003}$~\AA\ and $0.061_{-0.003}^{+0.003}$~\AA, respectively. For the mean stack, $W_{r,300}$ and $W_{r,500}$ are  $0.120_{-0.005}^{+0.005}$~\AA\ and $0.149_{-0.006}^{+0.006}$~\AA, respectively. The quoted errors on the $W_r$s are obtained from the 68\% confidence intervals of the $W_r$ distribution of 1000 bootstrap realizations. The measured $W_r$s indicate that \lya\ absorption is detected around our galaxy sample with $>99\%$ confidence interval for both the median and mean stacks, yet the absorption strength is very weak. The significant difference between the mean and median stack is due to skewed flux distribution. This difference is discussed in detail in \citet[]{Muzahid_2021}. Unlike their work, the distribution of pixel flux remains left-skewed for a velocity window of $\pm100$ \kms, giving rise to a higher median flux value for both the pseudo-continuum and absorption centroid compared to the mean, resulting in a smaller $W_r$. For all the subsequent analyses, we used the median stacked spectra.

In the right panel of Fig.~\ref{fig:all stack}, we show the 2D median optical depth map for the complete sample.
 At the low optical depth end, the color scale saturates to a dark-blue color which represents the median Lya optical depth of random regions (median \lya\ optical depth within $\pm$2000 \kms\ of 4595 random redshifts). A more detailed analysis of the optical depth map is presented in Section~\ref{od result}. An excess in optical depth compared to random regions is clearly evident out to $\approx 8$~pMpc LOS Hubble distance ($\approx600$~\kms\   assuming pure Hubble flow at the median redshift of $0.1$), consistent with the width of the median stacked flux in the left panel. The elongation of the excess optical depth along the LOS direction is the manifestation of redshift space distortion.

\subsection{ {\texorpdfstring{\lya}{}} rest frame equivalent width profile} 
\label{br dep}

One of our primary goals is to measure the \HI\ \lya\ equivalent width profile to understand how the cool, neutral gas is distributed in and around galaxies. With this motivation, we divided the impact parameter range of our sample (6--3000 pkpc) into 7 bins and produced stacks of \lya\  absorption. In order to sample both the inner and outer regions of the CGM, the first two bins are confined within 6-50 pkpc and 50-100 pkpc, respectively. The remaining $b$ range is divided into 5 logarithmic bins. The rest-frame \lya\ equivalent widths measured from the median stacked spectra are plotted against the median of the impact parameter bins in the left panel of Fig.~\ref{fig:b stack}. The red points denote $W_{r,500}$ and the blue points denote $W_{r,300}$. The error bars along the y-axis represent 68\% confidence intervals of the median $W_r$ distribution obtained from 1000 bootstrap realizations. The error bars along the x-axis represent 68\% confidence intervals of the impact parameter distribution in each bin. The median stellar mass and redshift of each impact parameter bin are indicated by the legends at the bottom of the plots.

 \begin{figure*}
\centering
    \begin{subfigure}{0.50\textwidth}
        \centering

         \includegraphics[width=1.0\linewidth]{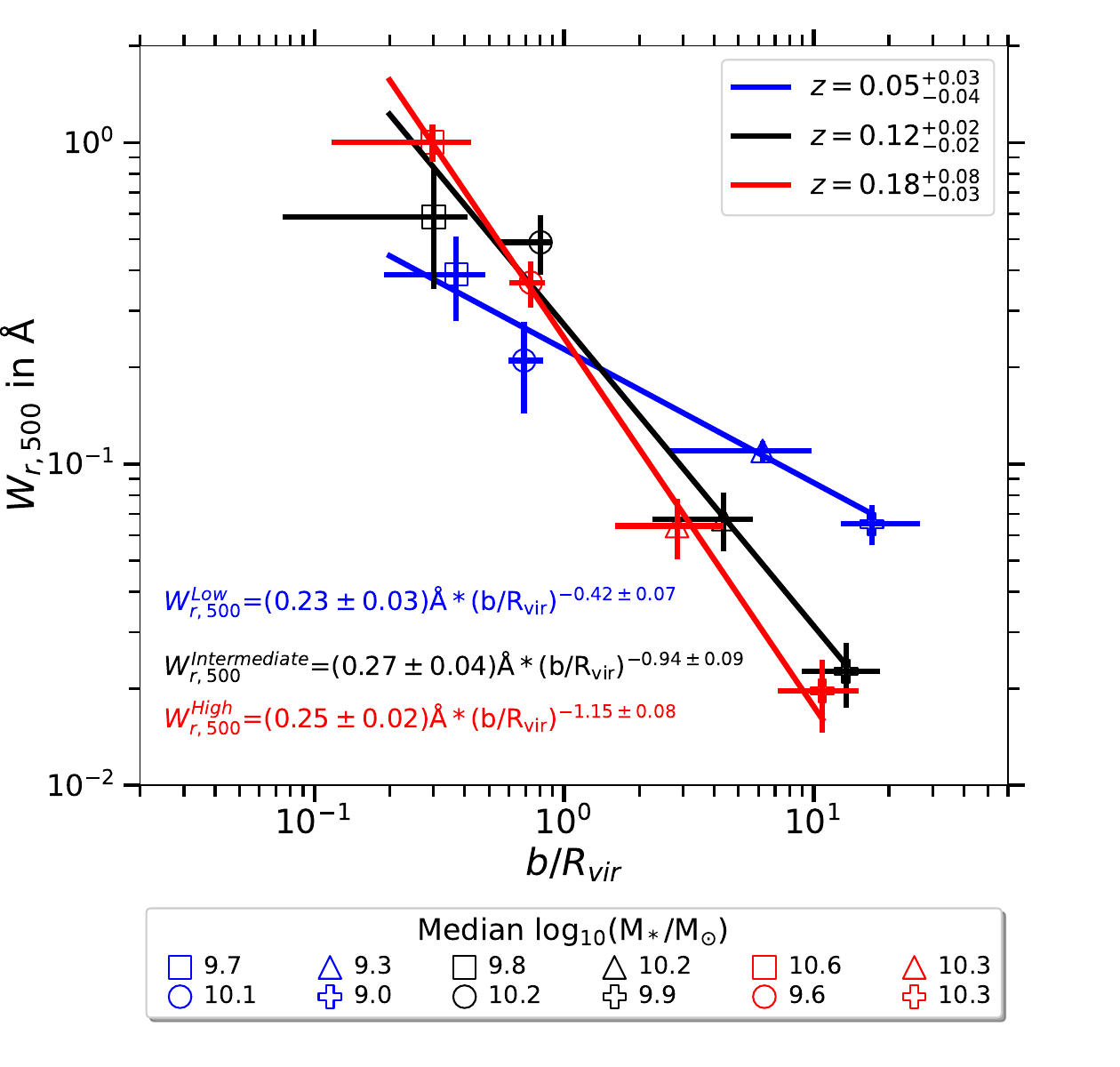}  
    \end{subfigure}%
    \begin{subfigure}{0.50\textwidth}
        \centering
   
        \includegraphics[width=1.0\linewidth]{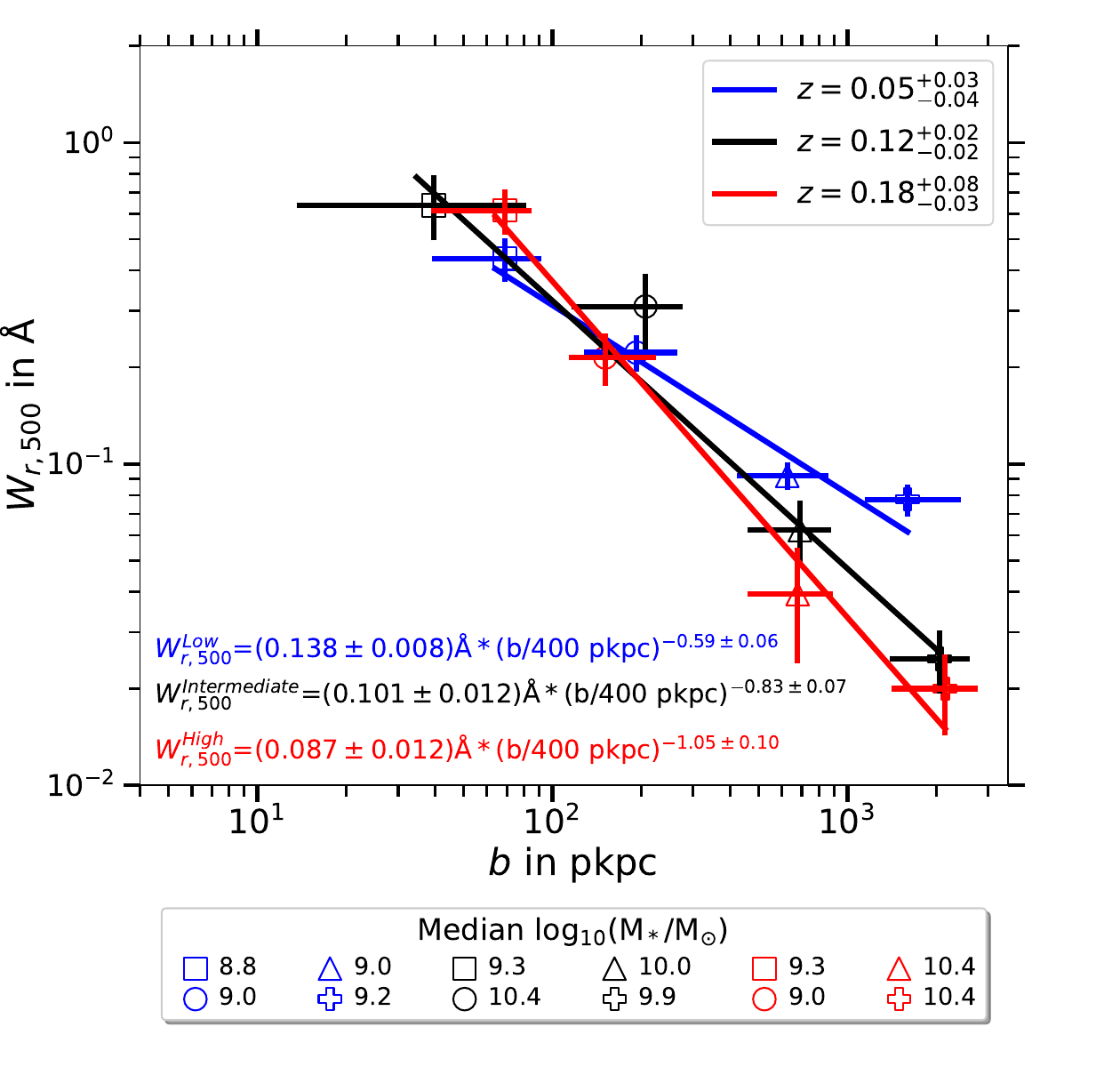}        

    \end{subfigure}
\caption{Dependence of the median \lya\ $W_{r,500}$-profile on redshift. The galaxy sample is divided into three tertiles of the redshift distribution indicated by the blue, black, and red  colors which correspond to median redshifts of 0.05, 0.12 and 0.18 as shown in the legends along with the 68\% confidence intervals. The data points represent median $W_{r,500}$ measurements plotted against median $b/R_{\rm vir}$ (left) and median $b$ (right) for each redshift bin. The median stellar masses of galaxies in each bin are listed in the legends below the plot. The error bars are similar to Fig.~\ref{fig:b stack}} 
\label{fig:z-b stack}
\end{figure*}

We have a large dynamic range in stellar mass for our galaxy sample. Empirical studies at low redshift as well as theoretical models have shown that the halo mass, as well as the halo size, increases with the stellar mass of galaxy. We, therefore, generated stacked \lya\ equivalent width profiles as a function of normalized impact parameter ($b/R_{\rm vir}$; see the right panel of Fig.~\ref{fig:b stack}). Instead of dividing the whole $b/R_{\rm vir}$ range in logarithmic bins, we created 3 bins for $b/R_{\rm vir}<1$ with roughly equal number of galaxies in each, and divided the remaining $b/R_{\rm vir}$ range (upto 25 $R_{\rm vir}$) in 4 logarithmic bins. The median stellar mass of each bin are indicated in the legends.

The median \lya\ $W_r$ shows a monotonic decline with both the impact parameter and normalized impact parameter. We find that a single power-law can adequately describe the data points. The best-fitting power-law relations are: 
\begin{align}
\label{eq2}
\begin{split}
W_{r,500}&=(0.117\pm0.005) \rm \angstrom \times(b/400~pkpc)^{-0.75\pm0.04} \\
W_{r,300}&=(0.106\pm0.004) \rm \angstrom \times(b/400~pkpc)^{-0.79\pm0.03}
\end{split}
\end{align}
and 
\begin{align}
\label{eq3}
\begin{split}
W_{r,500}&=(0.26\pm0.01)  \rm \angstrom\times(b/R_{\rm vir})^{-0.75\pm0.04} \\
W_{r,300}&=(0.25\pm0.01) \rm \angstrom\times(b/R_{\rm vir})^{-0.80\pm0.03} . 
\end{split}
\end{align} 
A fixed pivot near the middle of the impact parameter distribution at 400~pkpc is used for the $W_r$-profile as a function of the impact parameter to reduce the correlation of errors in slope and normalization. No such pivot is needed for the normalized impact parameter because $\rm log_{10}(b/R_{\rm vir})=0$ is near the middle of the $b/R_{\rm vir}$ distribution. Note that the power law indices in Eq.~\ref{eq2} and Eq.~\ref{eq3} are consistent with each other within 1$\sigma$, and that an excess \lya\ absorption around galaxy redshifts is detected with $>99\%$ confidence interval out to $\approx2$~pMpc (or equivalently $\approx 15 R_{\rm vir}$) in the transverse direction.

Although a single power-law can adequately explain the $W_r$-profile, it is to be noted that the second point in the left panel of Fig.~\ref{fig:b stack} is barely consistent with the model. In Section~\ref{lit_comp} we revisit the $W_r$-profile for the full sample where we explored the possibility of a power-law + log-linear (or Gaussian) model for the profile. All the measurements related to the $W_r$-profiles presented in this work are given as "online only" tables (see Appendix~\ref{online_only}).

 \begin{figure*}
\centering
    \begin{subfigure}{0.50\textwidth}
        \centering
  
        \includegraphics[width=1.0\linewidth]{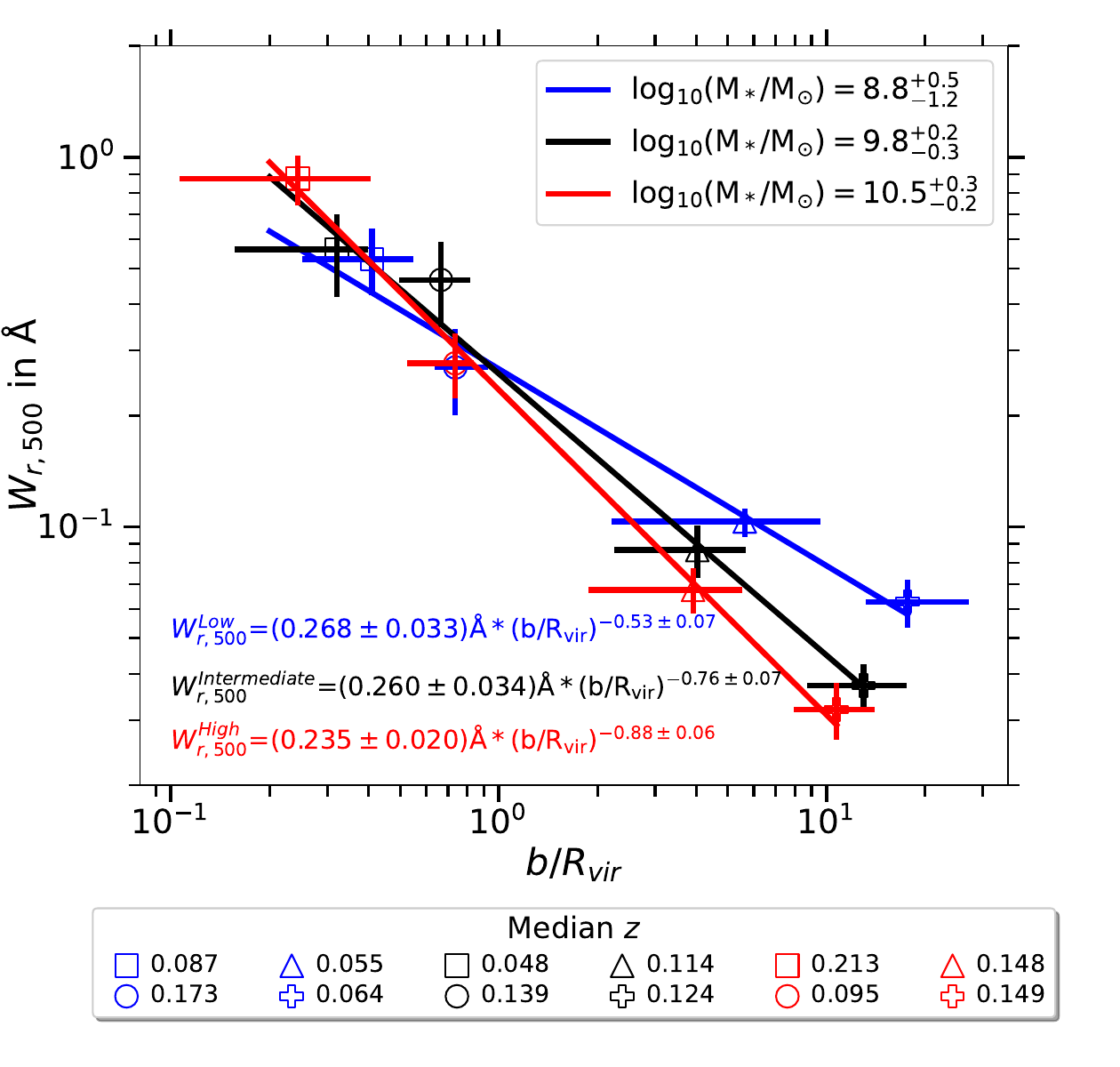} 
    \end{subfigure}%
    \begin{subfigure}{0.50\textwidth}
        \centering
      
        \includegraphics[width=1.0\linewidth]{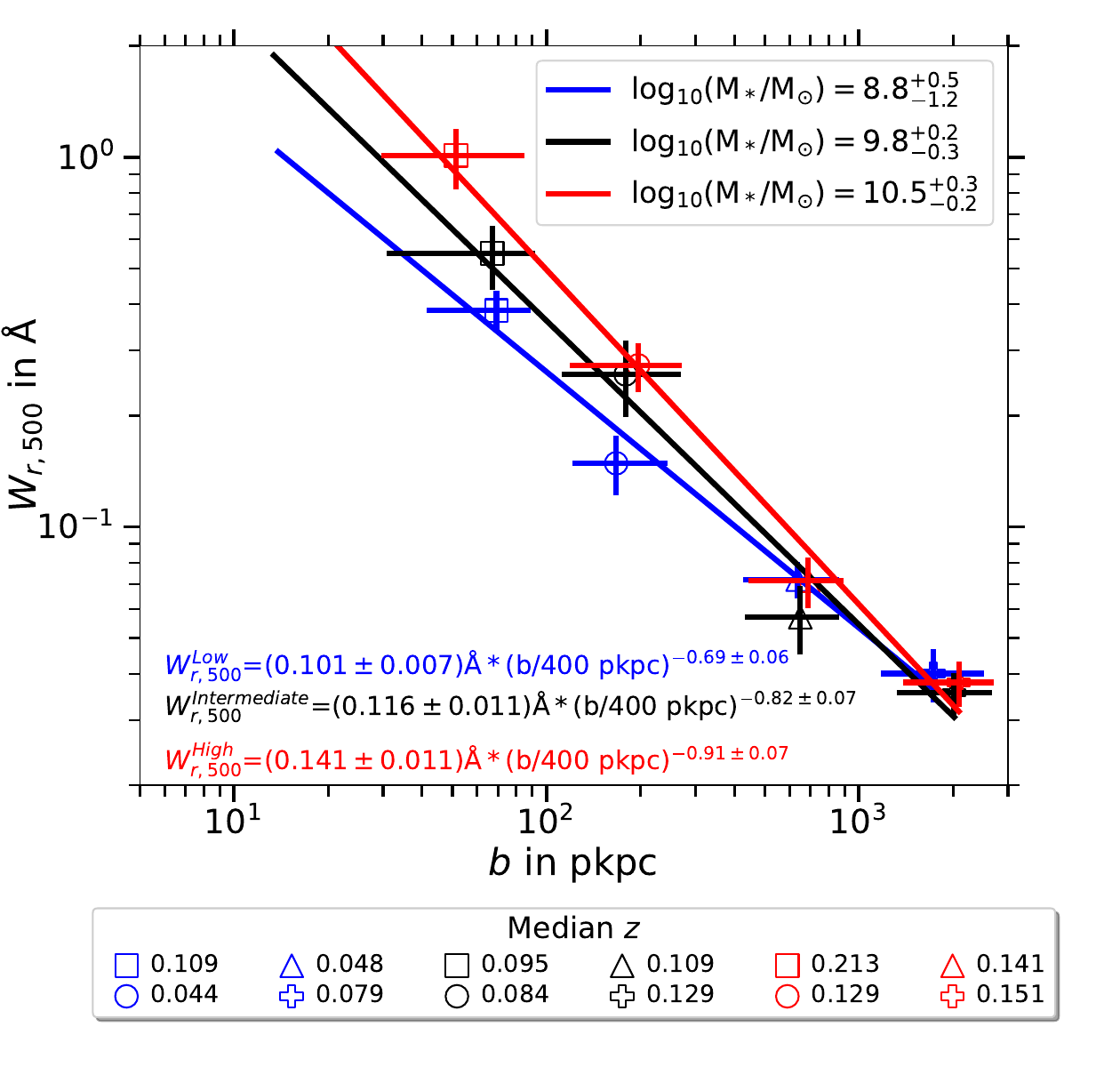} 
    \end{subfigure}
\caption{Dependence of the median \lya\ $W_{r,500}$-profile on stellar mass. The galaxy sample is divided into three tertiles of the stellar mass distribution indicated by the blue, black, and red colors. Median redshifts of galaxies in each bin are listed in the legends below the plot. The other details are similar to Fig.~\ref{fig:z-b stack}. The low-mass galaxy sample (blue) shows a significantly shallower slope as compared to the intermediate- (black) and high-mass (red) samples, particularly in the left panel.}
\label{fig:m-b stack}
\end{figure*}

\subsubsection{Variation with redshift}
\label{z dep}

The galaxies contributing to the \lya\ stack span a redshift range of $0.01-0.48$, corresponding to $\approx5$ Gyr of cosmic time. Since this is almost 35\% of the age of the universe, it is interesting to investigate whether the \lya\ $W_{r,500}$-profile evolves with time.

Fig.~\ref{fig:z-b stack} shows the redshift evolution of the median \lya\ $W_{r,500}$-profile. First, we split the galaxy sample into three tertiles of the redshift distribution, indicated by the blue (low-$z$, median $z=0.05$), black (intermediate-$z$, median $z=0.12$), and red (high-$z$, median $z=0.18$) colors. Each subsample is then further divided into 4 bins of impact parameters and normalized impact parameters. The first two normalized impact parameter bins are within the virial radius with roughly equal number of galaxies in each.  The last two bins outside the virial radius are logarithmic with a bin size of 1.0 dex, 0.8 dex, and 0.7 dex for the low-, intermediate-, and high-mass galaxies, respectively (see the left panel of Fig.~\ref{fig:z-b stack}). For the impact parameter, the first bin extends to 100 pkpc. The rest of the impact parameter range is divided into 4 logarithmic bins with a bin size of $\approx$0.5 dex for all three mass bins (right panel of Fig.~\ref{fig:z-b stack}). The median $W_{r,500}$ measured from the stacked spectra are plotted against the median values for $b/R_{\rm vir}$ and $b$ in each bin. Median stellar masses of each bin are listed in the legends below the plot.

In the left panel of Fig \ref{fig:z-b stack}, the low-$z$ points appear to lie beneath the intermediate- and high-$z$ points for $b < R_{\rm vir}$. This trend is reversed for $b/R_{\rm vir} > 1$, where the low-$z$ points lie well above the other points. A similar trend is also seen in the right panel of the figure, with the $W_{r,500}$-profile for the low-$z$ sample showing a shallower slope compared to the intermediate- and high-$z$ samples. This is supported by the best-fitting power-law relations indicated in the plots. The power-law indices for the intermediate- and high-$z$ bins are consistent with each other in both the left and right panels. However, the low-$z$ sample shows a significantly different (shallower) power-law slope. We point out that except for the second $b/R_{\rm vir}$ bin, the median stellar mass of the low-$z$ sample is almost an order of magnitude lower compared to the high-$z$ sample. This is not unexpected, since we already noticed a strong trend between redshift and stellar mass in Section~\ref{subsec:galprop} (see Table~\ref{tab:table_corr}). As such, the low-$z$ subsample has predominantly low-mass galaxies and that may be one of the reasons for the apparent difference. Further, the environment of these galaxies can give rise to the apparent redshift evolution. To mitigate the effects of this apparent redshift evolution, we control the redshift for further analysis whenever necessary.

\subsubsection{Variation with stellar mass}
\label{m dep}

The galaxies in our sample span a large range in stellar mass. The dependence of \lya\ absorption strength on  stellar mass at a given (normalized) impact parameter can shed light on the structural variation and self-similarity of the CGM. In order to investigate the possible mass dependence of the \lya\ $W_{r,500}$-profile, we proceed with a similar binning procedure as described in Section~\ref{z dep}.

Fig.~\ref{fig:m-b stack} shows the dependence of the \lya\ $W_{r,500}$-profile on stellar mass as a function of $b/R_{\rm vir}$ and $b$ in the left and right panels, respectively. The blue, black, and red colors represent, respectively, the low- (median \logm\ $=8.8$), intermediate- (median \logm\ $=9.8$), and high-mass (median \logm\ $=10.5$) galaxy samples. In the left panel of Fig.~\ref{fig:m-b stack}, the low-mass points are consistently above the intermediate- and high-mass points for $b>R_{\rm vir}$, indicating stronger \lya\ absorption around low-mass galaxies  outside the virial radii. However, no clear trend is seen for $b < R_{\rm vir}$. Overall, the \lya\ $W_{r,500}$-profile is significantly shallower for the low-mass sample (slope~$=-0.55\pm0.07$) compared to the intermediate-mass (slope~$=-0.77\pm0.07$) and high-mass (slope~$=-0.89\pm0.06$) samples.

In the right panel of Fig.~\ref{fig:m-b stack}, the low-mass galaxies show suppressed \lya\ absorption compared to their high- and intermediate-mass counterparts for a given impact parameter for the two inner-most bins. However, no significant difference in $W_r$ is seen among the three different mass samples for $b> 500$~kpc. Here we also notice a shallower slope for the low-mass sample, but the power-law indices are not as significantly different as in the left panel.

The presence of a strong correlation between $z$ and $M_*$ in our complete sample (Table~\ref{tab:table_corr}) is reflected in the median redshifts of the three mass bins. The redshifts in the legends show a general trend that the median $z$ is higher for the high mass subsample at a given $b$ or $b/R_{\rm vir}$ bin (with the exception of the second $b$ and $b/R_{\rm vir}$ bin). However, by controlling the redshifts of the galaxy samples, we confirmed that the trend of $W_{r,500}$ with $M_*$ is not driven by the underlying $z-M_*$ correlation (see Section~\ref{sec: mdep}).

 \begin{figure*}
\centering
    \begin{subfigure}{0.50\textwidth}
        \centering
      
        \includegraphics[width=1.0\linewidth]{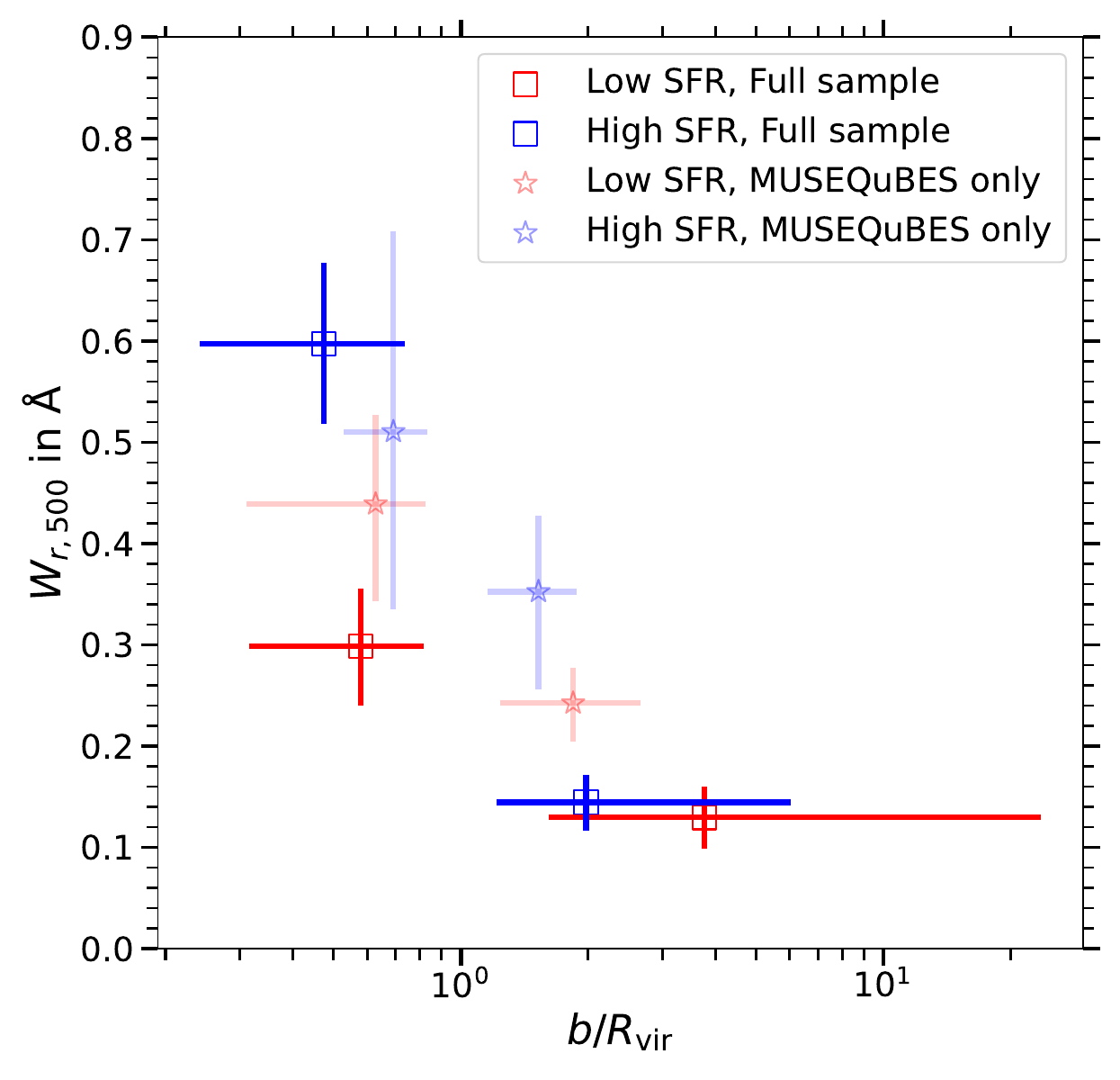}       

    \end{subfigure}%
    \begin{subfigure}{0.50\textwidth}
        \centering

        \includegraphics[width=1.0\linewidth]{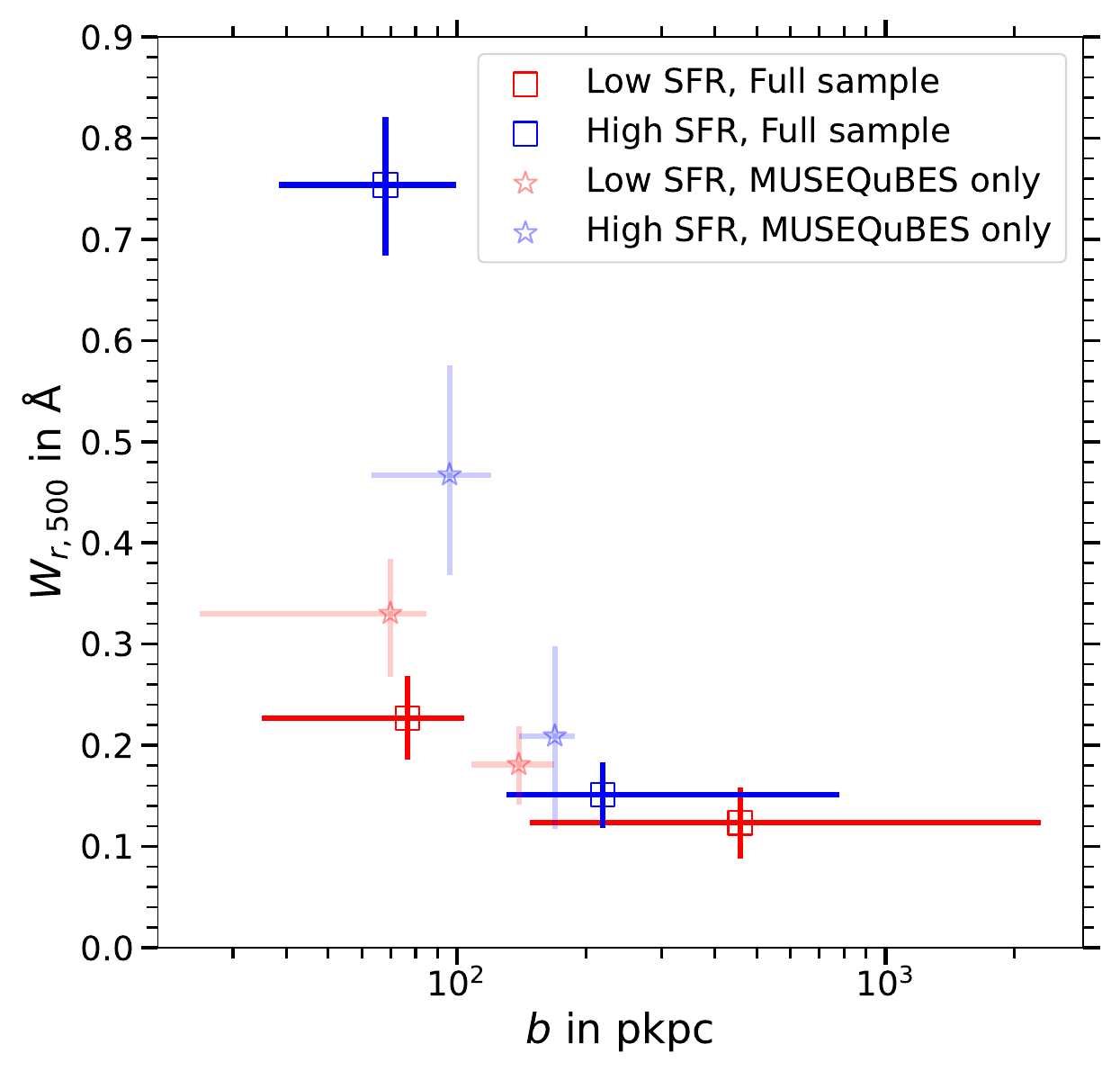}       

    \end{subfigure}
\caption{Dependence of the median \lya\ $W_{r,500}$-profile on SFR as a function of $b/R_{\rm vir}$ (left) and $b$ (right). Only two SFR bins and two $b$ (and $b/R_{\rm vir}$) sub-bins are created for stacking due to the reduced number of galaxies with SFR estimates. The red and blue open squares denote the $W_{r,500}$ measurements for the low- and high-SFR bins, respectively. The median SFR and redshift (along with the 68\% confidence interval) of each bin are tabulated in appendix (S7 and S8). Upper limits are treated as detections to compute the median SFR. The open stars in lighter shades indicate measurements for the MUSEQuBES subsample only. Error bars are similar to Fig.~\ref{fig:b stack}. A strong SFR dependence of the circumgalactic \lya\ absorption is seen only within the virial radius.}  
\label{fig:sfr-b stack}
\end{figure*}

 \begin{figure*}
\centering
    \begin{subfigure}{0.50\textwidth}
        \centering
     
        \includegraphics[width=1.0\linewidth]{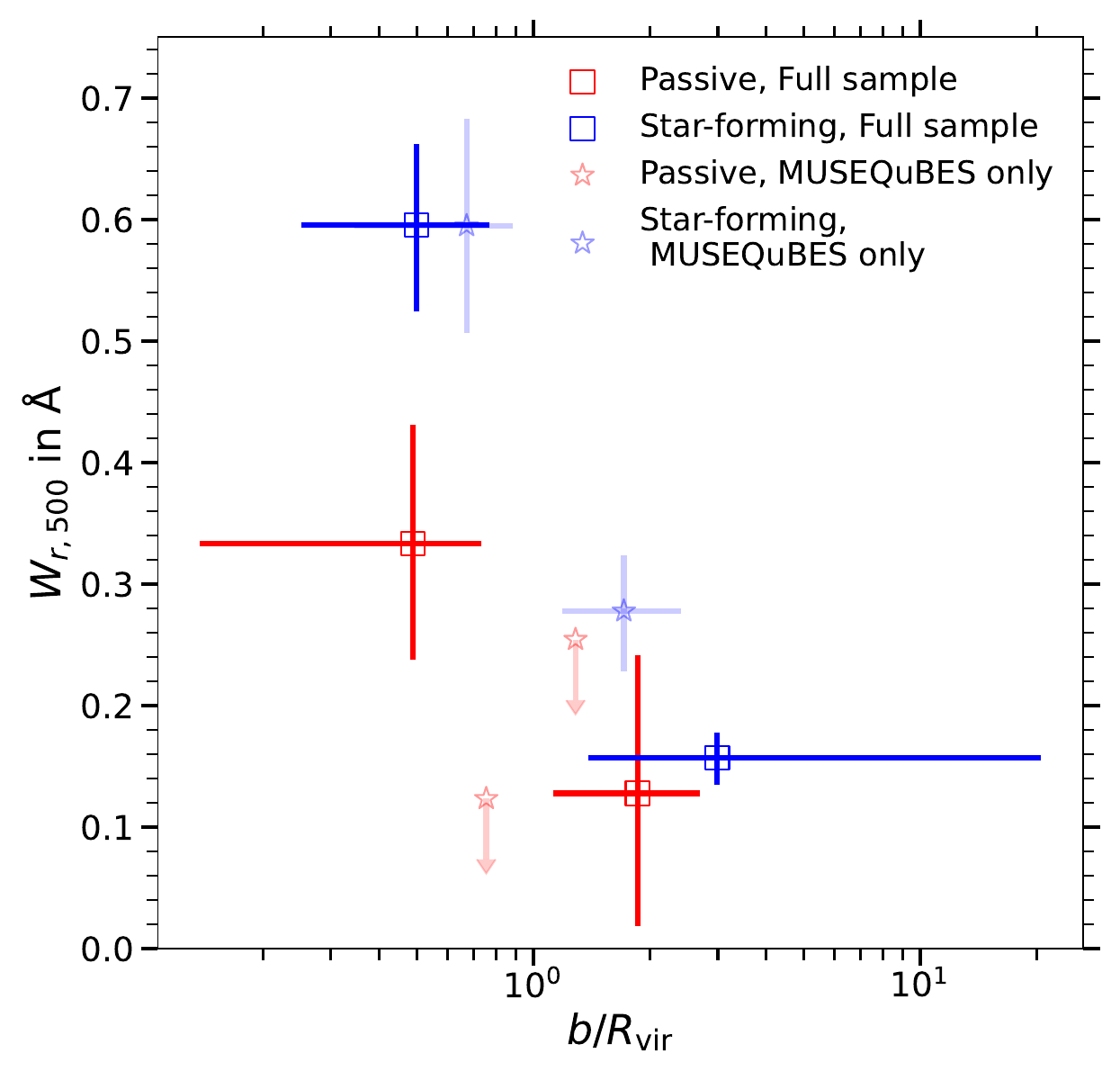}  
    \end{subfigure}%
    \begin{subfigure}{0.50\textwidth}
        \centering

        \includegraphics[width=1.0\linewidth]{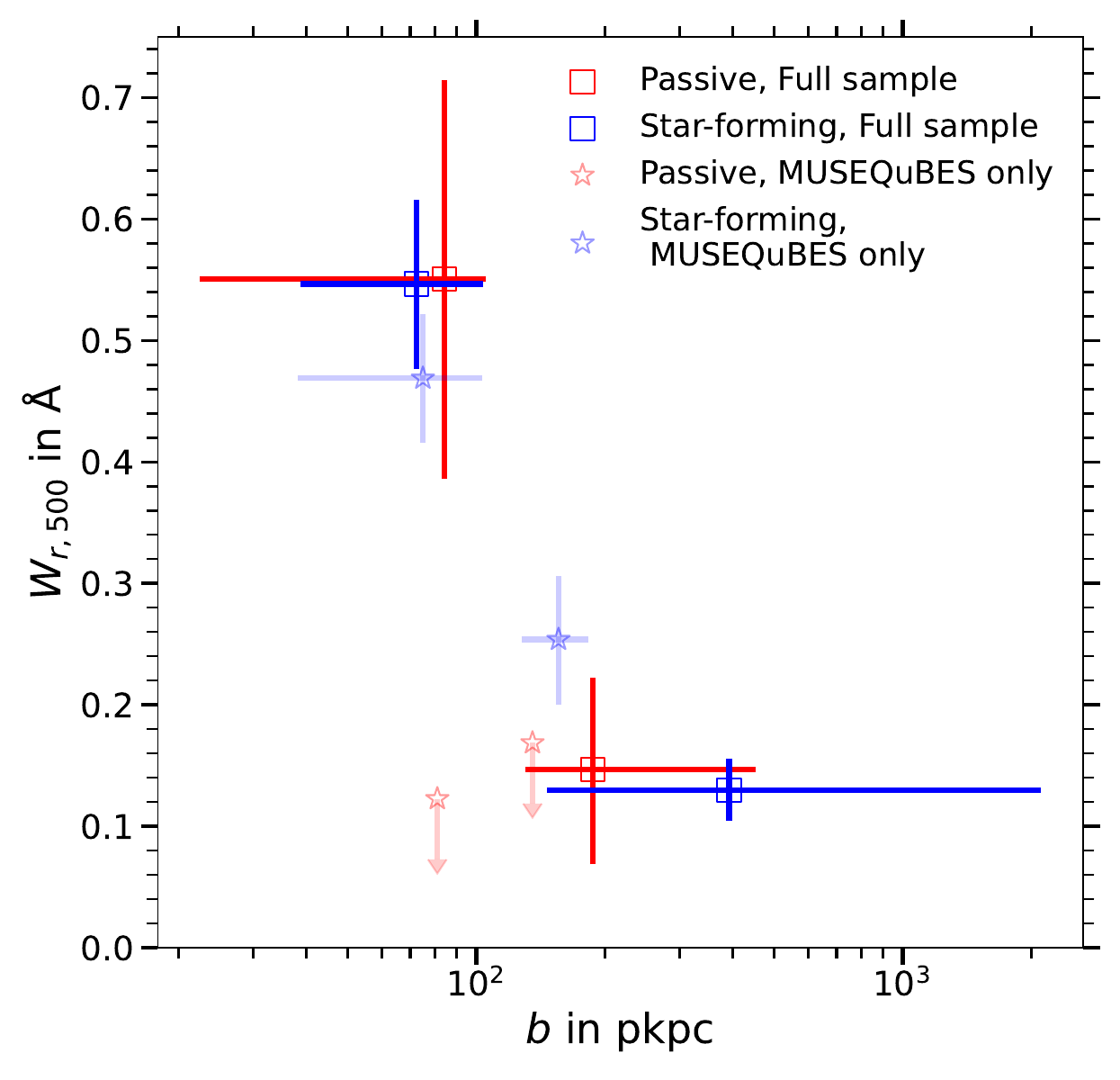}

    \end{subfigure}
\caption{Same as Fig.~\ref{fig:sfr-b stack} but for sSFR. A strong dependence of the median \lya\ $W_{r,500}$ profile on the sSFR is seen within the virial radius in the left panel. }  
\label{fig:ssfr-b stack}
\end{figure*}

\subsubsection{Variation with SFR}  
\label{subsec:SFRdep} 

The SFR of a galaxy is intimately related to the availability of cool gas in the ISM which is fuelled by the CGM. Star formation driven outflows expel metal-rich gas from galaxies to the CGM and/or IGM. A fraction of the metal-enriched gas may be recycled back to the galaxies aiding in further star formation. Therefore, connecting the CGM properties and SFR is imperative to understand the role of diffuse circumgalactic gas in galaxy evolution.

We used the 442 quasar-galaxy pairs comprising of 339 galaxies with measured SFR (including upper limits) to investigate the effects of the SFR on the \lya\ profile. The left and right panels of Fig.~\ref{fig:sfr-b stack} show the dependence on SFR of the $W_{r,500}$-profile plotted against $b/R_{\rm vir}$ and $b$, respectively. The data points in red and blue open squares represent the $W_r$ measurements for the low- and high-SFR bins. The median SFRs of the galaxies in individual bins are tabulated in Appendix~B (see Tables~S7 and S8).

The 442 pairs with SFR measurements are divided into two SFR bins based on the median SFR. The galaxies with measured SFR and upper limits below the median SFR are included in the low-SFR bin (median $10^{-1.6}~\rm M_{\odot}~\rm yr^{-1}$. Upper limits are treated as detection to compute the median). No galaxies with upper limits on SFR are included in the high-SFR bin (median $10^{-0.4}~\rm M_{\odot}~\rm yr^{-1}$). Galaxies in each SFR bin are further divided into two $b$ or $b/R_{\rm vir}$ bins. For $b/R_{\rm vir}$, the inner bin contains galaxies with $b/R_{\rm vir}<1$ and the outer bin contains galaxies with $b/R_{\rm vir} \ge 1$. For impact parameter $b$, we adopted a separating value corresponding to the 33 percentile of the $b$ distribution (119 pkpc and 107 pkpc for low- and high-SFR bin, respectively). In this way, the outer bin will have twice as many galaxies as the inner bin. This binning strategy is adopted to increase the signal-to-noise ratio of the stacked spectra for galaxies with large impact parameters for which the \lya\ absorption signal is intrinsically weak.

The high-SFR galaxies show significantly stronger \lya\ absorption for $b/R_{\rm vir} < 1$ as indicated by the blue triangle in the left panel of Fig.~\ref{fig:sfr-b stack}. A similar trend is also seen in the right panel for $b \lesssim 100$~pkpc. However, there is no clear difference in \lya\ absorption between the two SFR bins outside the virial radius (or at $>100$~pkpc). Although we have not explicitly controlled for redshift, the median redshift in a given $b$ or $b/R_{\rm vir}$ bin is similar for the high- and low-SFR galaxies (the values are tabulated in Appendix Table S7 and S8, along with the 68\% confidence intervals).

A similar exercise only with the MUSEQuBES galaxies gives rise to consistent results (open crosses with lighter shades). However,  the larger error bars, owing to the smaller numbers of galaxies contributing to the stacks, reduce the statistical significance of the results. The suppression of \lya\ $W_r$  outside virial radius for the full sample compared to MUSEQuBES only sample (for both high- and low-SFR bins) in the left panel of Fig.~\ref{fig:sfr-b stack} is due to the presence of more high impact parameter quasar-galaxy pairs in the full sample. This is evident from the median and the 68\% confidence interval of the $b/R_{\rm vir}$ and $b$ distributions shown in Fig.~\ref{fig:sfr-b stack}. The enhanced \lya\ absorption for the full sample within $\approx100$~pkpc is due to the enhanced median SFR of the contributing galaxies in this bin compared to the MUSEQuBES only galaxy sample ($\rm log_{10}(SFR/M_{\odot}~yr^{-1})=-0.2$ for the full sample as opposed to $-0.7$ for the MUSEQuBES only sample (See Table~S7 in Appendix~B)).

A similar analysis only for the MUSEQuBES galaxies is not carried out in Section~\ref{z dep} \& Section~\ref{m dep} owing to the inadequate number of pairs.

\subsubsection{Variation with sSFR}  
\label{subsec:sSFRdep}  

Most of the galaxies with SFR estimates in our sample are scattered around the star-forming main sequence relation (see Fig.~\ref{fig:prop}). The sSFR is thus a better diagnostic of star formation activity which is independent of stellar mass.

We investigate the dependence on sSFR by dividing the sample into star-forming and passive subsamples using a threshold sSFR of $10^{-11}~\rm yr^{-1}$. Each of these subsamples is again divided into two $b$ and $b/R_{\rm vir}$ bins following the same binning procedure as used in Section~\ref{subsec:SFRdep}. The left panel of Fig.~\ref{fig:ssfr-b stack} shows the median $W_{r,500}$-profile for star-forming and passive galaxies with blue and red open boxes, respectively, plotted against $b/R_{\rm vir}$. The right panel of Fig \ref{fig:ssfr-b stack} shows the same but plotted against $b$. While the median $W_r$-profiles for the two subsamples do not show any significant difference when plotted against $b$ in the right panel, the left panel reveals significantly stronger \lya\ absorption for the star-forming subsample for $b<R_{\rm vir}$. Nonetheless, no significant difference is seen outside the virial radius. Similar to Fig.~\ref{fig:sfr-b stack}, the redshifts of star-forming and passive galaxies in a given $b$ or $b/R_{\rm vir}$ bin are consistent (see Tables~S9 and S10 in Appendix~B).

With open stars of lighter shades, we show the \lya\ $W_{r,500}$-profiles for the two sSFR subsamples using only the MUSEQuBES galaxies. Consistent with the full sample, a significantly stronger \lya\ absorption is observed for the star-forming galaxies for $b < R_{\rm vir}$. Moreover, only a marginal difference between the star-forming and passive subsamples is seen outside the virial radius.

 \begin{figure*}
     \centering
    \includegraphics[width=0.9\textwidth]{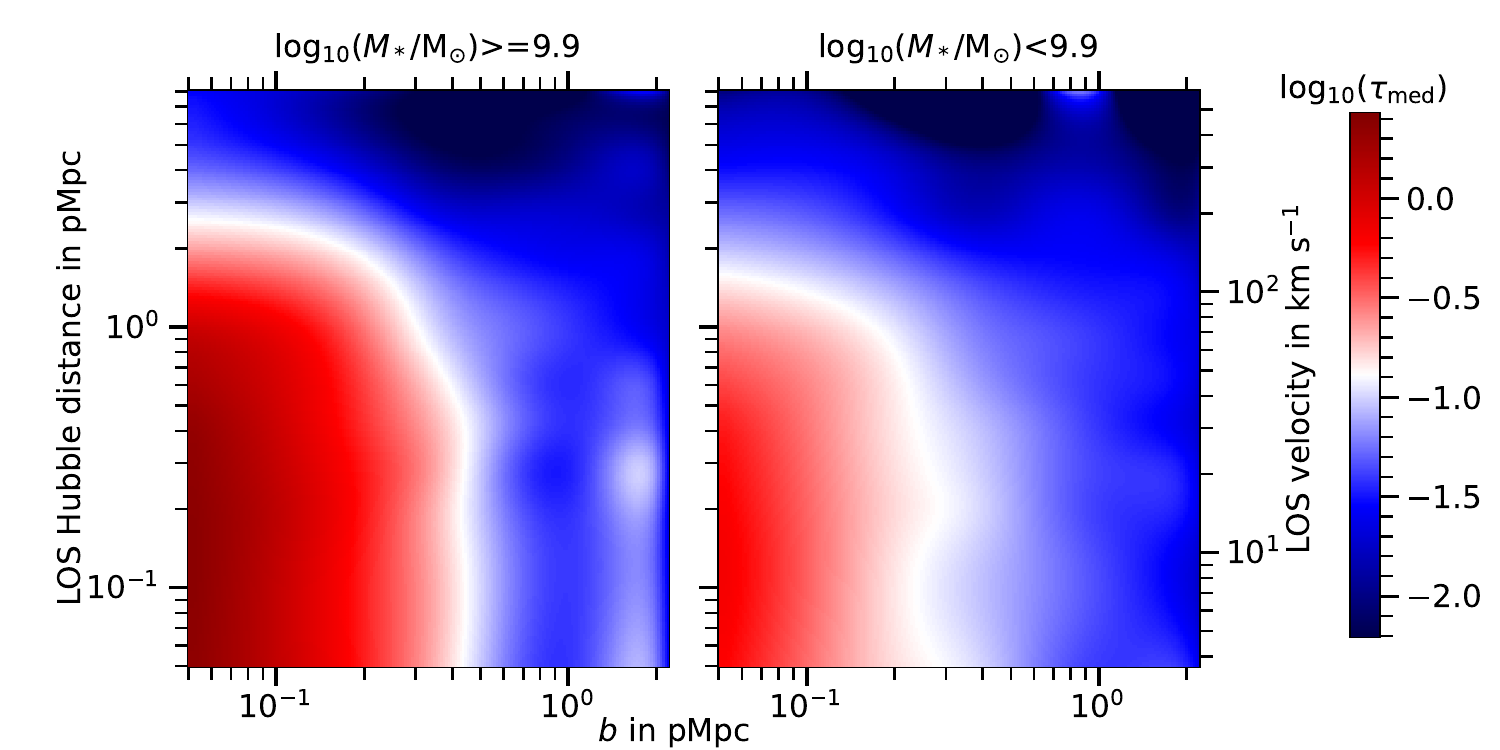}    
     \caption{2D median optical depth maps of \lya\ absorption around high-mass galaxies (left, median \logm\ $= 10.4$) and low-mass galaxies (right, median \logm\ $=9.2$). The two galaxy samples are  separated at ${\rm log_{10}}(M\rm_* / M_{\odot})=9.9$ , which is the median stellar mass of our sample. The color scale is common for both maps. 
     The median \lya\ optical depth is plotted as a function of impact parameter and LOS Hubble distance (LOS velocity in the right y-axis). Similar to Fig.~\ref{fig:all stack}, positive and negative velocity sides are folded to increase the $S/N$ ratio. The bin size and smoothing procedure are similar to Fig.~\ref{fig:all stack}. The high-mass sample shows a significantly stronger absorption signal in the inner parts of the map.}    
     \label{fig:OD map massbin}
 \end{figure*}

 The enhanced \lya\ absorption for the star-forming galaxies outside virial radius for the MUSEQuBES sample compared to the full sample in Fig.~\ref{fig:ssfr-b stack} can be attributed to the smaller values of $b/R_{\rm vir}$ and $b$.
The apparent inconsistency in results between the full and MUSEQuBES samples of passive galaxies in the smallest impact parameter bins of Fig.~\ref{fig:ssfr-b stack} is likely due to small number statistics. Only 6 and 7 galaxies contribute to the stack for the passive MUSEQuBES subsample in the two impact parameter bins (10 and 3 in the two normalized impact parameter bins). The 68\% confidence interval was not possible to compute for the small sample size in some MUSEQuBES bins (labelled with $\pm$0.00 range in the Tables in Appendix~B).

\subsection{Optical depth maps} 
\label{od result}

In order to inspect the correlation between galaxies and cool, neutral gas around them simultaneously along the transverse direction and along the line of sight direction, we produced 2D median optical depth maps following the procedure described in Section~\ref{od_recovery}.

The map for the complete sample is shown in the right panel of Fig.~\ref{fig:all stack}. The first LOS Hubble distance and impact parameter bin is constructed within 0.1 pMpc. Next, the bin size is 0.17 dex and 0.29 dex for the LOS Hubble distance and impact parameter, respectively. The LOS velocity around a galaxy is converted to LOS Hubble distance assuming pure Hubble flow at the galaxy redshift. The negative and positive LOS velocity differences are further merged to increase the $S/N$ in each bin. The optical depth for the random region is generated following the strategy described in Section~\ref{sec:results}. The median optical depth of all pixels from the random redshift stack represents the median random optical depth, which is set as the minimum of the optical depth color scale in the map. A Gaussian filter with half of the bin size is used to smooth the raw optical depth map.

 The right panel of Fig.~\ref{fig:all stack} reveals enhanced \lya\ optical depth compared to random regions out to 600 \kms\ or 8~pMpc along the LOS direction. The enhanced optical depth is observed out to 2~pMpc along the transverse direction. The apparent elongation of the excess optical depth in the LOS direction is reminiscent of the ``fingers of god'' effect and is reported earlier in the literature \citep[][]{Rakic_2012,Turner_14}. This is owing to the peculiar motions of the infalling and/or outflowing gas rather than redshift uncertainties \citep[see e.g.,][]{Rakic_2013,Turner_17, Chen_20}.

Next, we divided our sample into two stellar mass bins with \logm\ $=9.9$ as the separating value (i.e, the median stellar mass of the full quasar-galaxy pair sample). The median redshifts for the high- (median $10^{10.4}~\rm M_{\odot}$) and low-mass (median $10^{9.2}~\rm M_{\odot}$) galaxy subsamples are 0.145 and 0.088, respectively. However, we did not control  the redshift distribution here, as this reduces the number of available pairs leading to insufficient $S/N$. The OD maps for the high- and low-mass galaxy samples are  shown in the left and right panels of Fig.~\ref{fig:OD map massbin}, respectively. The median optical depth of random region for the high- and low-mass galaxy samples are $10^{-2.34}$ and $10^{-2.29}$, respectively. A common color scale is chosen for the maps which runs from the minimum of these two. We saturate the color scale at the maximum median optical depth of the two samples. The map for the high-mass galaxies shows a significantly stronger signal in the innermost transverse and LOS Hubble distance bin as compared to the low-mass sample. The excess optical depth around high-mass galaxies is also found to be more extended along the LOS direction compared to the low-mass counterparts  due to higher peculiar velocities, as predicted by simulations \citep[see e.g.,][]{Kim_2008, Rakic_2013, Turner_17}.

\begin{figure*}
\centering
        \centering        
        \includegraphics[width=0.9\linewidth]{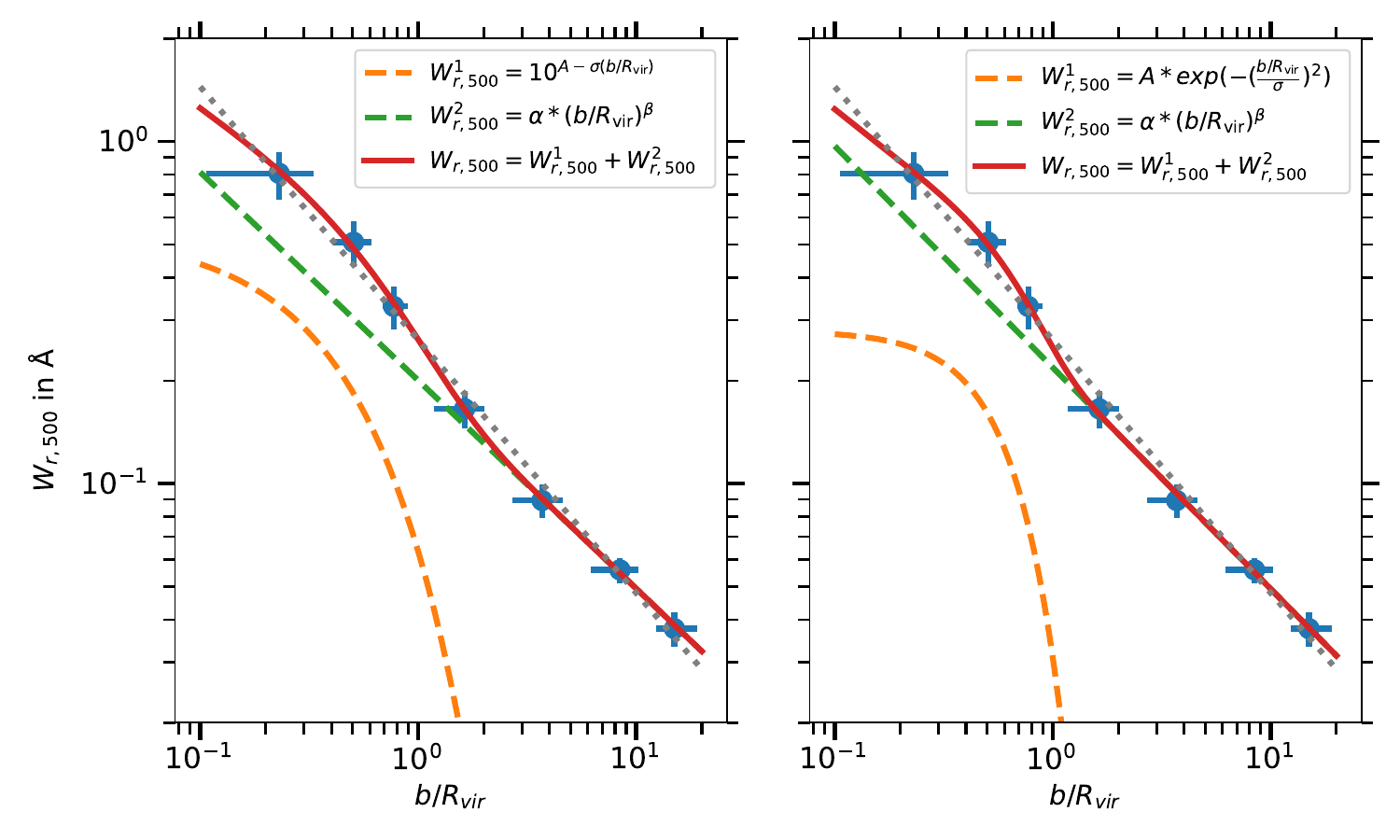}    
\vskip-0.5cm        
\caption{Modeling the observed \lya\ $W_{r,500}$-profile with two components. The data points show \rewf\ plotted against $b/R_{\rm vir}$ from this work. We use a log-linear (left, orange dashed) or a gaussian (right, orange dashed) and a power-law component (green dashed, both panels) as shown by the legends. The solid red line is the combined best-fit model in both panels.  The single-component power-law from Fig.~\ref{fig:b stack} is overlayed in both panels with grey dotted lines.} 
\label{fig:lit fit}
\end{figure*} 

\section{Discussion}
\label{disc}

 In this section, we discuss the main findings from Section~\ref{sec:results}.

\subsection{Comparison of the {\texorpdfstring{\lya\ $W_r$}{}} -profile with the literature} 
\label{lit_comp}

 We found a monotonically decreasing trend of the median $W_r$(\lya) with both impact parameter ($b$) and normalized impact parameter ($b/R_{\rm vir}$) (Fig.~\ref{fig:b stack}). Such a trend has been reported in the literature with individual \lya\ $W_r$ measurements in the CGM of low-$z$ galaxies \citep[see e.g.,][]{Prochaska_2011,Borthakur_2015} or from $W_r$ of mean stacked \lya\ absorption \citep[see e.g.,][]{Liang_14}. However, here we reported the presence of excess cool, neutral gas surrounding galaxies out to 2~pMpc, equivalently $\approx15R_{\rm vir}$. A single power-law with a slope of $\approx -0.75$ can adequately represent the median-stacked \lya\ equivalent width profile (Eq.~\ref{eq2} \& Eq.~\ref{eq3}). The single-component power-law index is in good agreement with \citet{Prochaska_2011}, obtained from individual \lya\ $W_r$ measurements for sub-$L_*$ and dwarf galaxies at similar redshifts ($z<0.48$) for  $b < 1$~pMpc. Using an $F$-test we confirmed that a two-component power-law model is unnecessary. However, this does not necessarily discard the possible presence of a secondary component other than a power-law.

Previously, \citet[]{Borthakur_2016} combined individual \lya\ $W_r$  measurements in the CGM of galaxies from the COS-Halos and COS-GASS surveys and obtained a log-linear $W_{r,500}$-profile for $b/R_{\rm vir} \lesssim 2$ with a slope of $0.387\pm0.103$. The power-law nature of the $W_{r,500}-$profile at large distances likely arises from the galaxy-absorber correlation which is well described by a power-law \citep[]{Tejos_14}. While going from smaller to larger impact parameters, a transition from a dark matter halo dominated environment to the regime dominated by halo-halo clustering has been observed in the gas surface density profile \citep[see][]{Zhu_2014}. Recently, \citet[]{Wilde_2023} used a Gaussian 1-halo term (arising from the CGM) along with a power-law 2-halo contribution (which is essentially the contribution due to galaxy-absorber clustering) to explain the observed \lya\ covering fraction profile. The transition point between the 1-halo and 2-halo terms is defined as the extent of the CGM in their work.

In order to look for a similar transition region between 1-halo and 2-halo contributions, we fit a log-linear and a Gaussian profile along with the default power-law model to the observed $W_{r,500}-$profile plotted against $b/R_{\rm vir}$. The two models are as follows:

\begin{equation}
\label{eq:loglin}
    W_{r,500}=10^{A - \sigma (b/R_{\rm vir})} + \alpha~(b/R_{\rm vir})^{\beta}
\end{equation}
and,
\begin{equation}
\label{eq:gauss}
    W_{r,500}=A~e^{-(\frac{b/R_{\rm vir}}{\sigma})^2}+ \alpha~(b/R_{\rm vir})^{\beta}
\end{equation}

The left and right panels of Fig.~\ref{fig:lit fit} show the best-fit models with Eq.~\ref{eq:loglin} and Eq.~\ref{eq:gauss}, respectively. The parameters of the best-fit models are summarized in Table \ref{tab:2comp_params}. The slope for the log-linear relation we obtain is somewhat steeper than found by \citet{Borthakur_2016} but consistent with $2\sigma$. An $F-$test rejects the null hypothesis that a single component power-law is a better representation of the $W_{r,500}-$profile compared to either of  the two-component models with a $p$-value of $>0.96$. The 1-halo term, represented by the Gaussian or log-linear function, contributes negligibly to the profile at $b/R_{\rm vir} \gtrsim 1$ which closely resembles the CGM-scale ($R_{\rm CGM}$) in the recent study of \citet{Wilde_2023}.

\begin{table}
\caption{Best-fit parameters of multi-component fit to the $W_{r,500}-$profile}
\label{tab:2comp_params}
\centering
\begin{tabular}{lrlll}
\hline
Model      & $A$ & $\sigma$ & $\alpha$ & $\beta$ \\ \hline
Model-1 & $-0.24\pm0.22$  &  $0.94\pm0.37$   &    $0.20\pm0.04$   & $-0.61\pm0.10$     \\ 
Model-2 & $0.29\pm0.18$   &   $0.68\pm0.23$  &     $0.22\pm0.03$  & $-0.65\pm0.07$     \\ 
\hline \hskip0.1cm 
\end{tabular}
Note-- Model-1: Log-linear+power-law; Model-2: Gaussian+power-law   
\end{table}

Using a single sightline (towards 3C$~$273), \citet[]{Morris1993} found that \lya\ absorbers indeed cluster around galaxies, but with a smaller amplitude than galaxy-galaxy clustering. \citet[]{Chen_2005, Chen_09} found that strong \lya\ absorbers cluster around emission-line dominated galaxies with a clustering amplitude comparable to the autocorrelation amplitude of emission-line galaxies. They also concluded that weak \lya\ absorbers cluster weakly around galaxies. \citet{Tejos_14} further showed that $\approx50$\% of weak \lya\ lines are correlated with galaxies at large transverse distances. The low column density Ly$\alpha$ absorbers can be attributed to the gas present in filament-like structures. Consistent with this picture, \citet{Wakker_2015} found a trend of increasing \lya\ equivalent width and line width with decreasing filament impact parameter. The \lya\ detection rate is $\approx80$\% within 500~pkpc of galaxy filament. Recently, \citet[]{Bouma2021} studied the relation between \lya\ absorbers and nearby galaxy filaments. They found an excess incidence rate (d$N$/d$z$) near filaments and a somewhat shallower slope for the column density distribution function compared to the general population of \lya\ absorbers at $z \approx 0$. They also noted that the strongest \lya\ absorbers are preferentially detected near galaxies or filament axes, albeit with a significant scatter. The observed excess \lya\ $W_r$ far outside the virial radii of galaxies can, in part,   come from the CGM of other galaxies (`intra-halo' gas) residing in the filaments. The `inter-halo' gas tracing the underlying density fields of the filaments can also contribute. The $W_{r,500}-$profile in our fit is dominated by a  single component power-law outside the virial radius. Such a power-law is widely usually used in the context of galaxy-galaxy clustering. However, the power-law component is regarded as the 2-halo term in our work. Although, strictly speaking, the gas constituting the 2-halo component is expected to have an `intra-halo' origin, we emphasize that this can account for gas in large-scale structures or filaments that are not part of the CGM of any (detectable) galaxy.

The connection between \lya\ absorbers and galaxies, over a scale of $>500~h^{-1}\rm kpc$, is investigated by \citet{Dave_99} using hydrodynamical simulations. They obtained a power-law index of $-0.71$ for the $W_{r,500}$-profile around low redshift galaxies, which is in good agreement with our single component power law fit. To investigate the origin of the $W_{r,500}-b$ correlation in their simulations, they determined the power-law index of the $W_r$-profile for the diffuse gas phase with densities and temperatures typical of the IGM. The overall power-law index and the index for the diffuse gas phase were found to be statistically indistinguishable. Based on this finding, they argued that the trend between $W_r$ and impact parameter can arise from the clustering of gas and galaxies over large-scales. In the next section, we will examine whether the stacked \lya\ profile can be explained by the large-scale clustering of gas and galaxies.

\begin{figure*}
  \hskip -5.1cm 
  \hbox{\vbox{
      \includegraphics[width=0.518\textwidth]{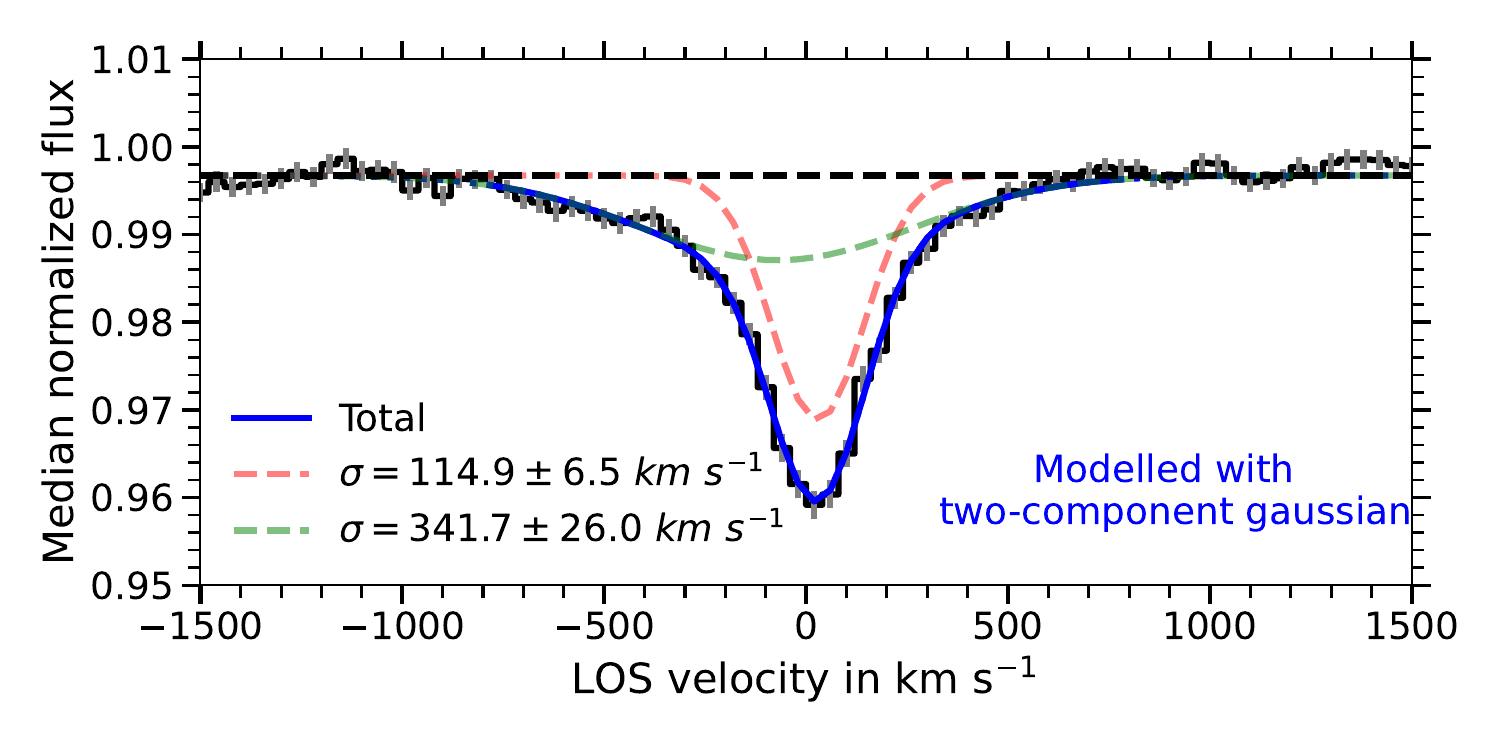}      
      \includegraphics[width=0.5\textwidth]{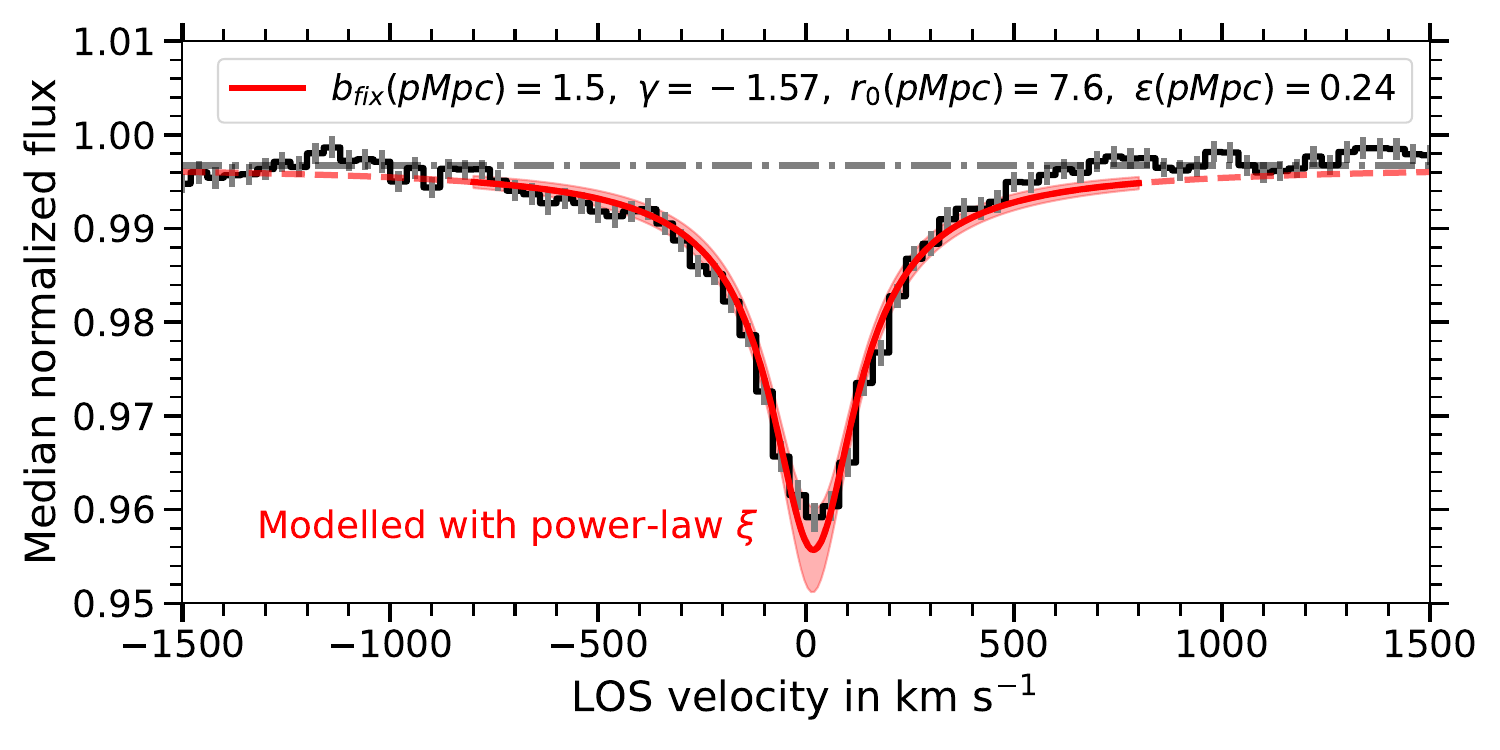}
 }
 \hskip -4cm 
      \includegraphics[width=0.5\textwidth]{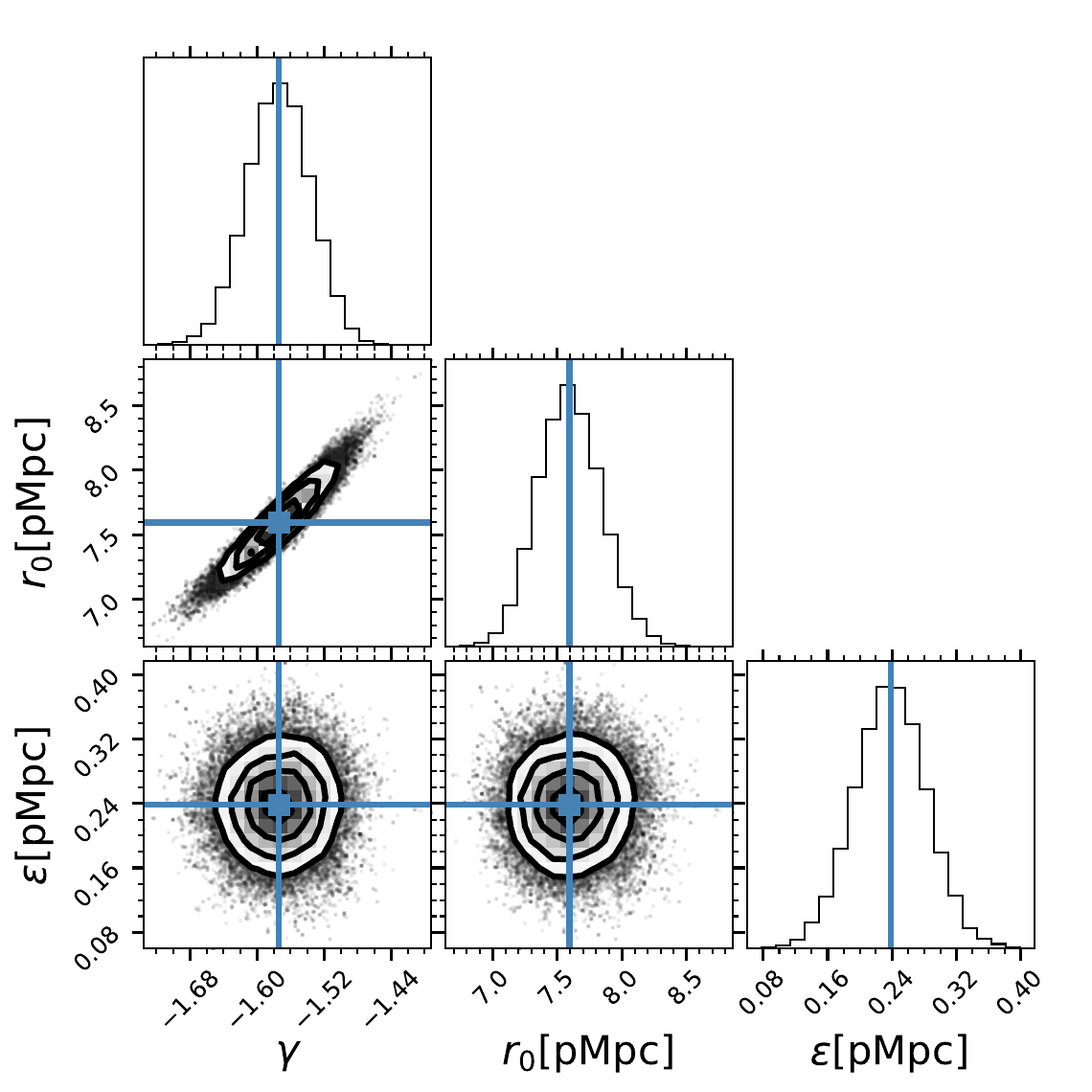}
}
\caption{{\tt Left-top:} The observed median stacked \lya\ absorption profile for the complete sample (black histogram). The blue solid line indicates the two-component Gaussian fit to the observed profile (extrapolated region shown with dotted line), with the individual components shown in red and green dashed lines. The standard deviations of the individual Gaussian components are indicated in the legends. The grey error bars are 68\% confidence intervals of the median flux in each bin coming from 1000 bootstrap realizations. {\tt Left-bottom:} Observed profile fit with the 2-point correlation function given in Eq.~\ref{2pt_form} shown by the red solid line (extrapolated region shown with red dashed line). The shaded region shows 1$\sigma$ scatter around the best-fit relation. {\tt Right:} Corner plots showing one and two-dimensional projections of the posterior probability distributions of the parameters used to fit the median stacked profile.} 
\label{fig:width_sim}
\end{figure*}

\subsection{The role of galaxy-absorber clustering} 

The stacked spectral profile for the complete sample is mainly shaped by the velocity distribution of the \lya\ absorbers with respect to the host galaxies.  If the width of the velocity distribution is mainly determined by the random redshift errors, a Gaussian is a reasonable choice to model the stacked profile. However, the left-top panel of Fig.~\ref{fig:width_sim} shows that a single-component Gaussian cannot explain the observed median stacked flux profile of \lya\ absorption for the full sample. Two distinct Gaussian components with $\sigma$ values of $114.9\pm6.5$ and $341.7\pm26.0$~\kms\ are required to adequately explain the entire profile. The narrower and stronger Gaussian component likely arises from the \lya\ absorbers originating in galactic halos (CGM). The origin of the shallow and wider Gaussian component, however, can be more complex. It can arise from weak and broad individual \lya\ absorption associated with galactic halos (i.e., so-called broad \lya\ absorbers (BLAs) buried in the noise of individual spectra). Another possible origin is the narrow and loosely correlated \lya\ absorbers likely arising from the large-scale structures tracing the same overdensities as the galaxies. 

The large-scale clustering between galaxies and \lya\ absorbers has been characterized in the literature \citep[see e.g.,][]{Chen_2005,Tejos_14,Wilde_2021,Borthakur_2022}. The galaxy-absorber two-point correlation function can be parameterized as:
\begin{equation}
\label{eq:2pt_or}
  \xi(b, r_{\parallel})=\left(\frac{r}{r_0}\right)^{\gamma},  
\end{equation}
 where, $r_0$ and $\gamma$ are the scale-radius and power-law slope, respectively. The 3D distance $r = \sqrt{b^2 + r_{\parallel}^2}$. The correlation function $\xi(b, r_{\parallel}) $ is essentially a measure of the excess number of absorbers as compared to random regions, hence it can be written as \begin{equation}
     \xi = \frac{N_{\rm obs}-N_{\rm rand}}{N_{\rm rand}}; 
\end{equation} 
where, $N_{\rm obs}$ and $N_{\rm rand}$ are the number of observed and random absorbers at a given $b$ and $r_{\parallel}$.

At large impact parameters of $\sim1$ pMpc, the absorbing gas is unlikely to arise from individual galactic halos. As discussed in  Section~\ref{lit_comp}, the contributing absorbers most likely reside in large-scale structures tracing the same overdensities as galaxies. At this length scale, it may be safe to assume a similar distribution of absorption strength for the uncorrelated \lya\ absorbers and for those absorbers which are correlated with the galaxies. The excess \lya\ absorption we see at large impact parameter can be interpreted as the increased number of absorbers around galaxies due to clustering, and not necessarily because of increased strength. With this assumption, the two-point correlation function can be directly related to the observed median stacked flux.

The optical depth can be written as $\tau=\langle a\rangle N$, where $\langle a\rangle$ and $N$ are the average strength and the number of contributing absorbers, respectively. For the median stack, the median optical depth at a given LOS velocity bin $v^j$ can be written as $$\tau^j_{\rm med}=N_{\rm med}^j\langle a\rangle_{\rm med}^j~.$$ Therefore, $$\tau^j_{\rm med,obs}=N_{\rm med,obs}^j\langle a\rangle_{\rm med,obs}^j~,$$ and   $$\tau^j_{\rm med,rand}=N_{\rm med,rand}^j\langle a\rangle_{\rm med,rand}^j~.$$ The assumption of similar absorption strength distributions for the random and clustered absorbers implies $\langle a\rangle_{\rm med,obs} \approx \langle a\rangle_{\rm med,rand}$. Thus,  
\begin{equation}
   \frac{\tau^j_{\rm med,obs}- \tau^j_{\rm med,rand}}{\tau^j_{\rm med,rand}} = \frac{N^j_{\rm med,obs}- N^j_{\rm med,rand}}{N^j_{\rm med,rand}} \equiv \xi(b_{\rm med}, r_{\parallel}).
   \label{defzeta} 
\end{equation}
Note that the observed median stacked flux can be written as, 
\begin{equation}
     f_{\rm med,obs}= e^{-(\tau_{\rm med,obs}-\tau_{\rm med,rand} + \tau_{\rm med,rand})}~.
     \label{deffn} 
\end{equation} 

Combining Eq.~\ref{defzeta} and Eq.~\ref{deffn} and using $e^{-\tau_{\rm med,rand}}=f_{\rm med,rand}$, we obtain 
\begin{equation}
\label{f_pow}
    f_{\rm med, obs}^j=e^{-\tau_{\rm med, rand}\times (\xi(b_{\rm med}, r_{\parallel}^j)+1)}~. 
\end{equation} 
Here we dropped the superscript $j$ from $\tau_{\rm med,rand}$, as the optical depth of the random region does not depend on the LOS separation from galaxy redshift. 

We fix $b_{\rm med}$ to the median impact parameter of our galaxy sample of $1.5$~pMpc. The value of $\tau_{\rm med, rand}$ is obtained by converting the median flux at random regions  (essentially the pseudo-continuum of the median stack) to optical depth. To allow for a small velocity offset of the centroid of the stacked spectrum from $0$~\kms, we introduce a parameter $\epsilon$ in the two-point correlation function such that: 
\begin{equation}
\label{2pt_form}
  \xi(b, r_{\parallel},\epsilon)=\left(\frac{\sqrt{b^2+(r_{\parallel}-\epsilon)^2}}{r_0}\right)^{\gamma}~,   
\end{equation}
 where, we wave used the 3D distance $r=\sqrt{b^2+(r_{\parallel}-\epsilon)^2}$ in Eq.~\ref{eq:2pt_or}. The offset parameter $\epsilon$ is a purely mathematical construct to account for the velocity offset that can arise due to the finite LOS velocity bin size of 40~\kms.

Fitting the observed median stacked rest-frame \lya\ absorption profile with Eq.~\ref{f_pow} and Eq.~\ref{2pt_form} yields  best fit $r_0$ and $\gamma$ values of $7.6$~pMpc and $-1.57$, respectively. The best-fit  \footnote{We used the {\sc emcee} package \citep[]{emcee} of {\sc Python} for error estimation.} $\epsilon$ value is $0.24\pm 0.06$ pMpc, corresponding to $18\pm 4$~\kms\ which is well within the LOS velocity bin size used in our analysis.
The best-fitting model profile is shown in the left-bottom panel of Fig.~\ref{fig:width_sim} with the red solid line. The corner plot in the right panel of Fig.~\ref{fig:width_sim} shows the 1- and 2-dimensional projections of the posterior distribution for the model parameters. 

 The reduced-$\chi^2$ of $1.9$ obtained for this fit is larger than the reduced-$\chi^2$ of $0.6$ obtained for a double-component Gaussian fit. However, a fit with this simple two-point correlation function provides a significant improvement compared to a single component Gaussian fit (reduced-$\chi^2$ = 2.7). This indicates that the broad component in the stacked profile for large impact parameters can be partly explained by a simple two-point correlation function arising from the large-scale clustering between galaxies and \lya\ absorbers. 

At this point, we emphasize that previous studies on clustering analyses measured the $r_0$ and $\gamma$ from the correlation function projected along the line of sight, thus eliminating the effect of redshift space distortions. The correlation length-scale and power-law index obtained in our analysis however suffer from redshift space distortions since the two-point correlation function is not projected along the line of sight. The evidence for redshift space distortion is also clear from the optical depth maps (see Figs.~\ref{fig:all stack} \& \ref{fig:OD map massbin}). The correlation length of $7.6$~pMpc found in this work is considerably larger compared to the projected galaxy -- \HI\ absorber correlation length of $r_0=1.6h_{70}^{-1}$~Mpc found by \citet[]{Tejos_14}. This is also true even when we compare the $r_0$ by fixing $\gamma=-1.44$ as reported in \citet[]{Tejos_14}). Recently, \citet[]{Wilde_2021} presented projected correlation lengths of $3.4 h_{68}^{-1}$~Mpc and $6.2 h_{68}^{-1}$~Mpc for high ($M_*>10^{9.9}~\rm M_{\odot}$) and low ($M_*<10^{9.2}~\rm M_{\odot}$) mass galaxies at low redshift. The power-law index for the projected correlation function varies from $-1.2$ to $-1.9$ for low- to high-mass galaxies in their work.
Nonetheless, this exercise demonstrates that the presence of the significant non-Gaussian wings in the median stack can be interpreted as a consequence of weak, correlated absorbers likely arising from the large-scale structures tracing the same overdensities as the galaxies rather than warm/hot, widespread gas associated with individual halos.

\begin{figure}
    \centering
    \includegraphics[width=1.02\linewidth]{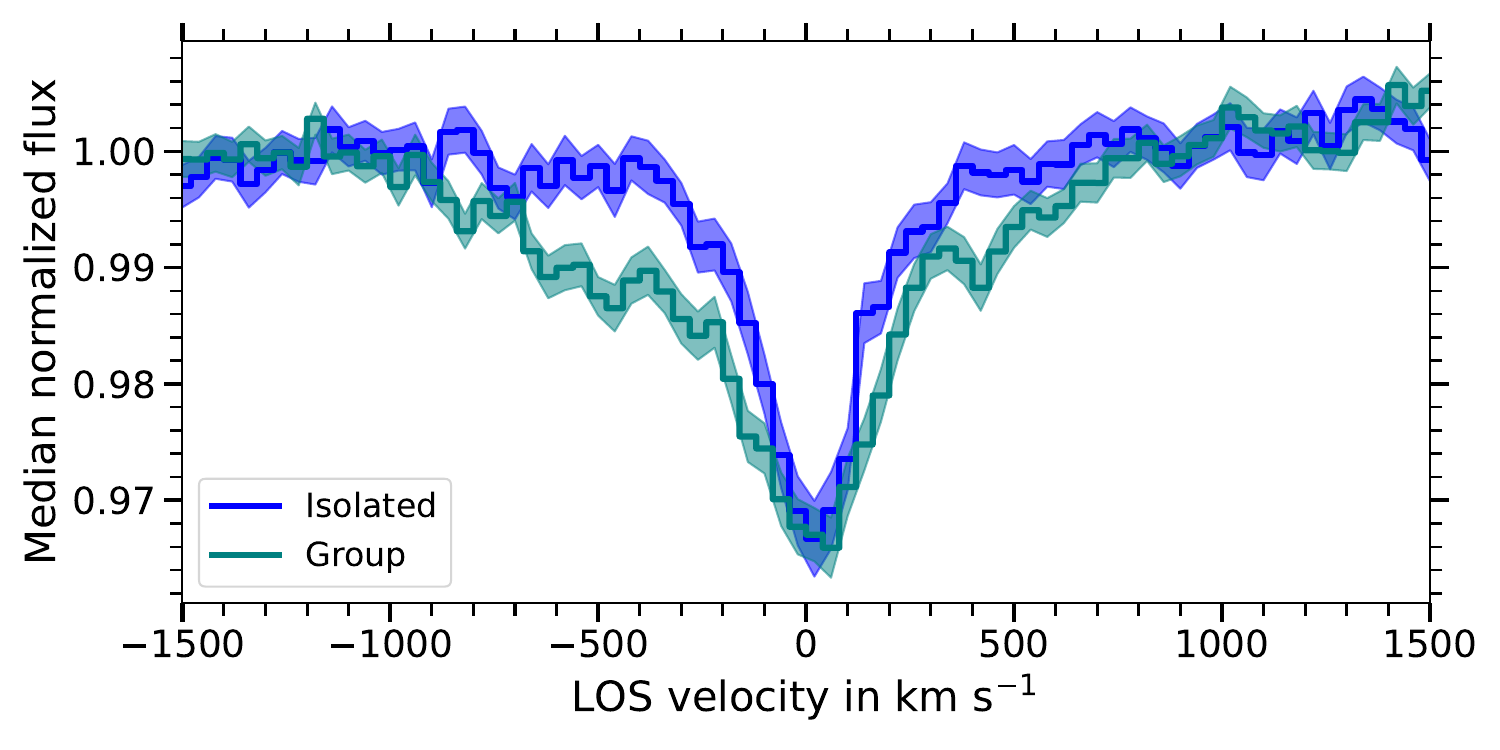}
    \caption{ The median stacked \lya\ absorption profiles for the isolated and group subsamples shown with magenta and teal histograms, respectively. The shaded region represents the 68\% confidence interval of median flux distributions obtained from 1000 bootstrap realizations. }
    \label{fig:env_comp}
\end{figure}

\begin{figure*}
      \centering
\includegraphics[width=0.42\linewidth]{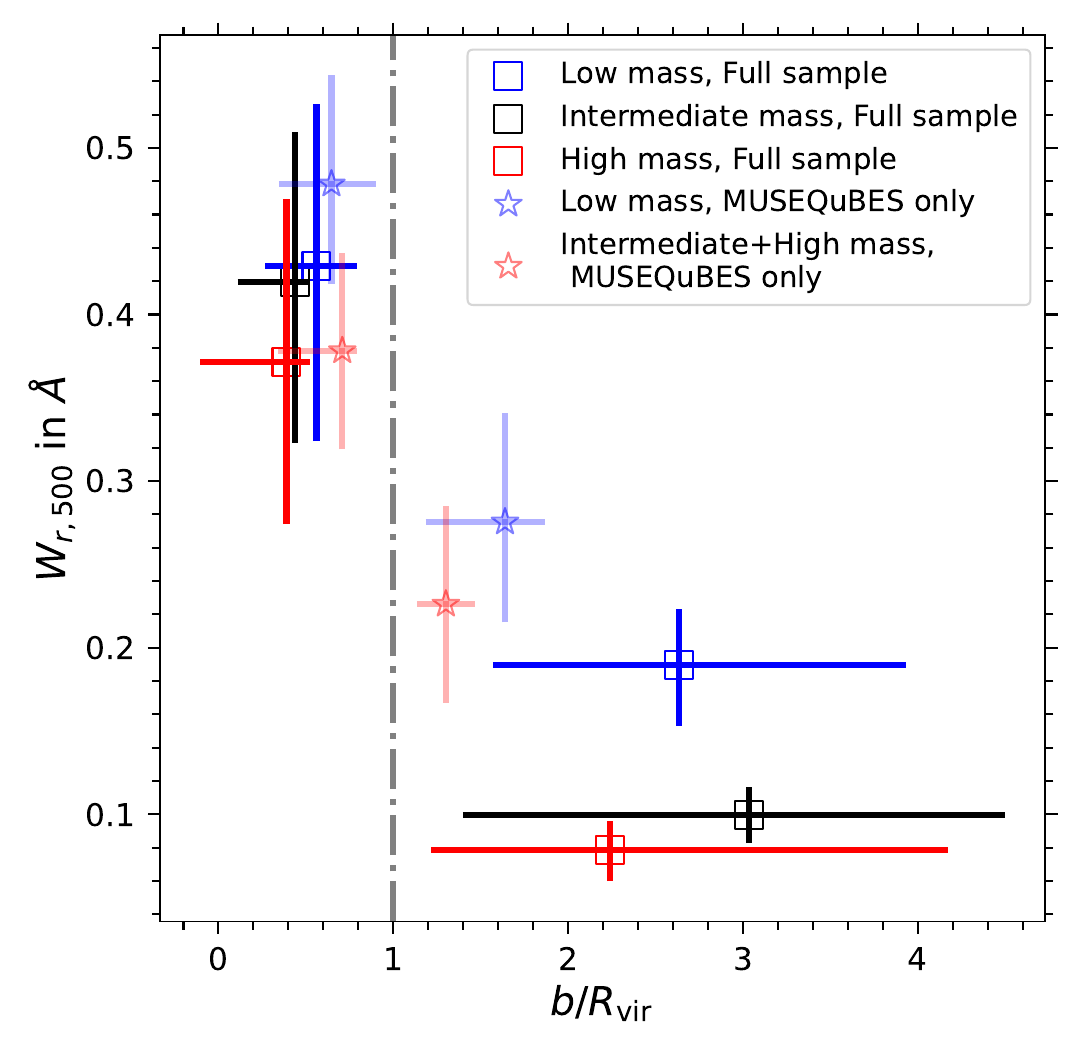}%
\includegraphics[width=0.62\linewidth]{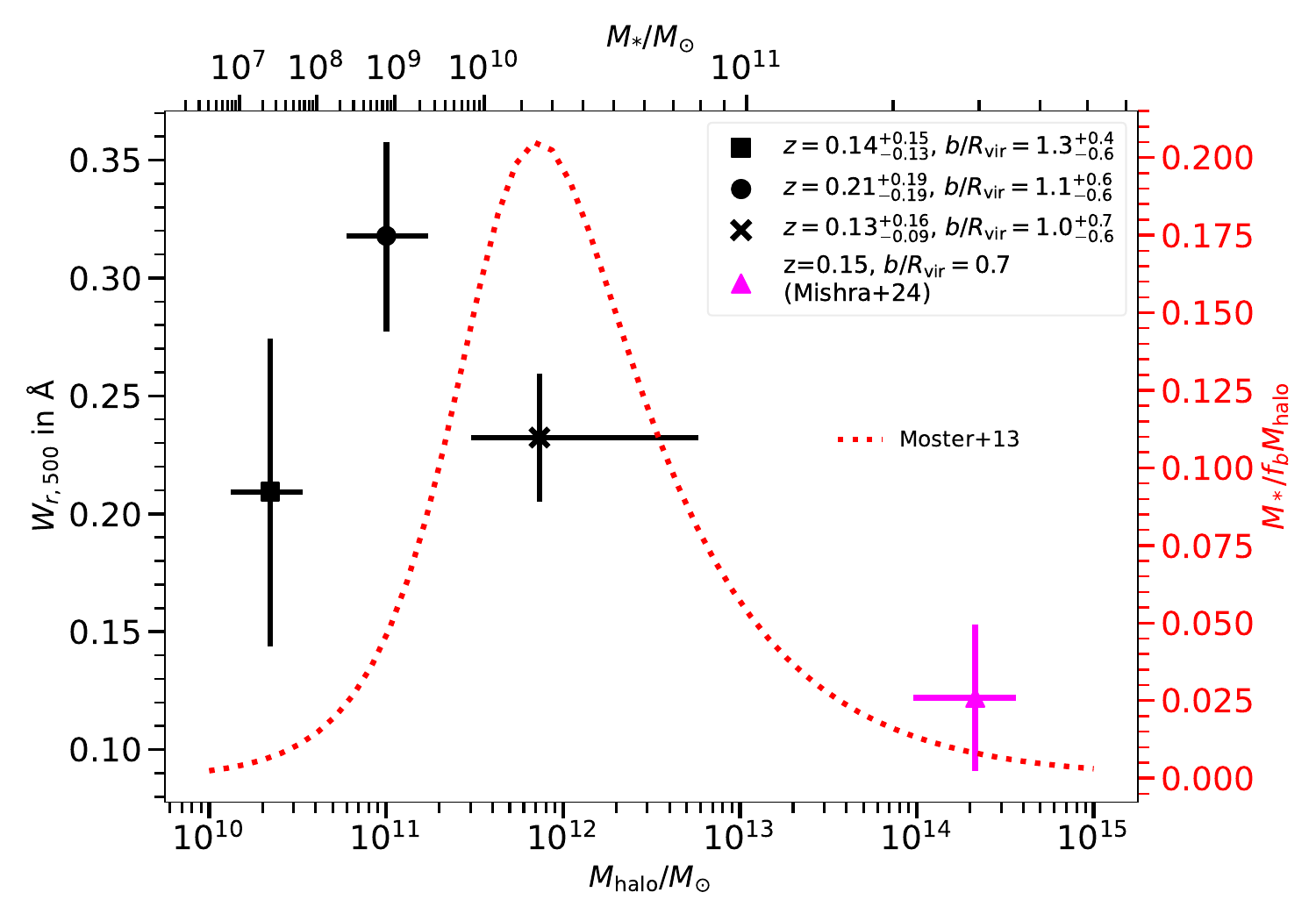} 
\caption{{\tt Left:} The effect of $M_*$ on the $W_{r,500}$-profile. The redshift is controlled for galaxies in a given $b/R_{\rm vir}$ bin. We first divided the galaxy sample into two bins with $b/R_{\rm vir}>1$ and $<1$. Galaxies in each $b/R_{\rm vir}$ bin are split by \logm\ $<9,~>10$, and $9<$\logm\ $<10$. Redshift-controlled galaxies are selected in each $M_*$ sample for a given $b/R_{\rm vir}$ bin to produce the \lya\ stacks. $W_{r,500}$ of median stacks are shown by blue, black, and red open squares for low-, intermediate-, and high-mass bins as a function of $b/R_{\rm vir}$. Measurements for the MUSEQuBES subsample only are shown with lighter shades and different markers  (one bin is made by combining intermediate- and high-mass galaxies due to the lack of \logm$>10$ galaxies in MUSEQuBES sample). 
 The data points inside the virial radius are plotted with offsets in the abscissa for clarity.
The error bars are similar to Fig \ref{fig:m-b stack}. {\tt Right:} The \lya\ $W_{r,500}$, measured within 2$R_{\rm vir}$, as a function of stellar/halo mass is plotted with black symbols. All galaxies with $0.25<b/R_{\rm vir}<2$ are selected and split into 3 stellar mass bins with \logm\ $<8,~>9.5$ and $8<$\logm\ $<9.5$ to produce the \lya\ stacks. The magenta triangle represents the median $W_{r,500}$ for stacked \lya\ absorption in low redshift ($z<0.5$) cluster outskirts ($b/R_{500}<1.5$) by  \citet[]{Mishra_2024}. The x-error bars show the 68\% range of $M_*$ in each bin, y-error bars are 68\% confidence intervals of the median $W_r$. The median $b/R_{\rm vir}$ and $z$ (along with 68\% confidence intervals) of each bin are given in the legends. The overlayed red dotted line showing the star-formation efficiency (see text) corresponds to the right y-axis.  } 
  \label{fig:m-b stack_cont}
 \end{figure*} 

 Another possible origin of the broader component could be the environment of the galaxies. We have employed a simple friends-of-friends (FoF) algorithm with a projected separation of $10'$ and LOS separation of $500$~\kms\ to identify overdense galactic environments.
  We have defined `groups' as the overdense region of more than four galaxies identified by the FoF algorithm with the aforementioned linking lengths.
 A power-law galaxy-galaxy clustering with a slope of $-1.8$  and correlation length of 3$h^{-1}$ Mpc is expected to give rise to $\approx4$ galaxies in the region of interest (at $z\approx 0.1$) due to clustering \citep[see][for the formalism]{Cherrey_2023}. The adopted number of five or more galaxies within the linking lengths thus suggests an overdense region. At this point, we emphasize that these overdensities are not necessarily virialized. 
 In Fig.~\ref{fig:env_comp}, we show the stacked \lya\ absorption profile for isolated and group galaxies with blue and teal histograms, with the shaded region denoting the corresponding 68\% confidence interval obtained from 1000 bootstrap realizations.
 The median stacked \lya\ absorption from the group galaxies showed symmetric secondary peaks at around $\approx \pm 500$~\kms, but no such feature was present in the stacked spectrum of isolated galaxies. While we acknowledge that it is challenging to characterize the environment of galaxies observed from different surveys with varying depths and completeness, the presence of the secondary peaks in the stacked spectrum for the group galaxies is consistent with the findings of \citet[]{Muzahid_2021} for a homogeneous sample of $z\approx3$ galaxies. Moreover, we obtained a consistent result when we considered the galaxies with $>90\%$ completeness from the Keeney+18 survey. A detailed analysis of the effects of the galaxy environment on the CGM will be presented elsewhere. However, we emphasize that the secondary peaks observed in the stacked spectrum of group galaxies are very different compared to the smooth wings of the observed spectrum, and they cannot be explained by a power-law cross-correlation function. The observed broad component in the stack at large impact parameters is more likely the manifestation of the galaxy-absorber correlation function.

\subsection{The role of stellar mass} 
\label{sec: mdep}

The observed $W_{r,500}$-profile is shallower for the low-mass galaxies than for their high-mass counterpart (see Fig.~\ref{fig:m-b stack}). This owes to the fact that the high-mass galaxies show stronger \lya\ absorption in the inner regions ($b<0.5R_{\rm vir}$) whereas the low-mass galaxy sample shows relatively stronger \lya\ absorption outside the virial radius. We note however that we did not control for the redshifts of the different mass bins. Consequently, the median redshift of the high-mass bin is higher compared to the low-mass bin.

In the left panel of Fig \ref{fig:m-b stack_cont}, we show the \lya\ $W_r$ measurements for three mass bins within and outside the virial radius. 
We first divided the galaxy sample into
two bins with $b/R_{\rm vir}>1$ and $\leq 1$. Galaxies in each $b/R_{\rm vir}$ bin are split by \logm $< 9$, $> 10$, and $9 <$\logm $< 10$. In each normalized impact parameter bin, we now controlled the redshifts of the galaxies so that the redshift distributions of the low-, intermediate-, and high-mass samples are similar.{\footnote{We chose equal number of galaxies within some narrow $\Delta z$ for the three mass bins. This ensures that the redshift distributions (and the median values) of the three mass bins are very similar.} The median \logm\ and median redshifts of different bins are tabulated in Appendix Table S11. Outside the virial radius, the \lya\ $W_r$ is significantly higher for the low-mass bin compared to the intermediate- and high-mass bins. Inside the virial radius, we notice a  decline in $W_r$ for the high-mass bin compared to the low- and intermediate-mass bins, but consistent within the $1\sigma$ uncertainties. The error bars are similar to Fig.~\ref{fig:m-b stack}. The data points inside the virial radius are plotted with small offsets in the abscissa for clarity. Measurements for the MUSEQuBES galaxies for a similar binning strategy are shown with a lighter shade and open star symbols. Similar to Fig.~\ref{fig:sfr-b stack}, the enhanced \lya\ absorption outside virial radius for MUSEQuBES galaxies compared to the full sample can be attributed to a lower median $b/R_{\rm vir}$ compared to the full sample. The trend for the MUSEQuBES galaxies is consistent with the full sample, although not significant due to a smaller sample size.

The presence of significant Ly$\alpha$ absorption outside the virial radius can arise from the extended CGM and/or the large-scale structures around galaxies. Cosmological hydrodynamical simulations by \citet{Voort_12} showed that the ``cold gas'', which can be traced by Ly$\alpha$, is mainly residing in filamentary structures around galaxies. Moreover, the geometry of such cosmic filaments is much more widespread (up to 4 $R_{\rm vir}$) for low-mass halos as compared to high-mass halos (see their figure~{1}), leading to higher gas covering fraction for low-mass galaxies outside virial radius. \citet[]{Johnson_17} reported the presence of \lya\ absorption well beyond the virial radius of dwarf galaxies. Recently, the observations of \citet[]{Wilde_2021} also showed that the \lya\ covering fraction remains roughly constant at $\approx 40\%$ out to $1$~pMpc from galaxies with \logm $<$9.2. The covering fraction for the high-mass galaxies, however, declines rapidly with impact parameter. Although their quoted $1\sigma$ errors have significant overlaps in different mass bins, this may hint at the steeper slope for the $W_r$-profile for the high-mass galaxies.

 A possible reason for the enhanced \lya\ absorption outside the virial radius for low-mass galaxies could be the environment. However, different completeness of the different surveys hinders a proper investigation of environmental effects for our complete sample. To circumvent this, we selected a subsample of galaxies from Keeney+18 with total completeness $>$90\% (see their Table 8). In that subsample, isolated galaxies are searched for using a simple 3D FoF algorithm with a linking separation of $10'$ \footnote{linking separation of 1 pMpc produces consistent result} and a linking velocity of 500 \kms. We found enhanced \lya\ absorption around low-mass galaxies outside the virial radius (median $b/R_{\rm vir} \approx 6$) even for this well-chosen sample of isolated galaxies. Therefore, we conclude that the observed mass dependence cannot be directly attributed to the environment. However, if a significant fraction of these low-mass galaxies is satellite, then their $R_{\rm vir}$ measurements can be underestimated. This can lead to the apparent excess `outside' the virial radius of the low-mass galaxies. Indeed, the convergence of the $W_{r,500}$ profile for the three mass bins at large $b$ (see the right panel of Fig.~\ref{fig:m-b stack}) implies that there is no excess \lya\ absorption at large proper distances around low-mass galaxies.  
 
 The underlying gas distribution in cosmic filaments can explain the origin of excess \lya\ $W_r$ far outside the virial radius for low-mass galaxies compared to high-mass galaxies when plotted against the normalized impact parameter. At a similar normalized impact parameter, massive galaxies trace a region further away in the filament compared to less massive or dwarf galaxies. This can cause the apparent enhancement of \lya\ $W_r$ outside virial radius when plotted against normalized impact parameter. However, the underlying gas distribution tracing the overdensity in the filament will produce similar absorption for both low- and high-mass galaxies at large physical distances,  consistent with our observation. In section \ref{lit_comp}, we suggested that the power-law 2-halo term outside the virial radius can account for the gas in the filaments of `inter-halo' origin as well.

A marginal suppression in the \lya\ absorption inside the virial radius for high-mass galaxies compared to the low-/intermediate-mass bin is observed when the redshift is controlled. To further explore this with larger galaxy samples, we selected all galaxies with $0.25<b/R_{\rm vir}<2.0$ in three mass bins with $\log_{\rm 10}(M_*/\rm M_{\odot})<8$, $8<\log_{10} (M_*/\rm M_{\odot})<9.5$, and $\log_{\rm 10}(M_*/\rm M_{\odot})>9.5$ to generate \lya\ stacks. The lower cut in $b/R_{\rm vir}$ is motivated by the lack of low-mass galaxies below this limit in our galaxy sample. Thus, a cut like this roughly brings the  $b/R_{\rm vir}$ distributions of the three mass bins on a similar footing. Although we did not explicitly control for the redshift, the median redshifts are comparable for the three mass bins (this is also true for the 68\% ranges).

The right panel of Fig.~\ref{fig:m-b stack_cont} shows the median \lya\ $W_{r,500}$ as a function of stellar mass (top axis)  and halo mass (bottom axis). To convert the stellar masses to halo masses, we used the sub-halo abundance matching relation of \citet{Moster_2013}. The median $W_{r,500}$ for our highest mass bin shows a clear suppression compared to the intermediate mass bin but is comparable to the low-mass bin.  The (pink) data point is obtained from \citet[]{Mishra_2024} which represents the median \lya\ $W_{r,500}$ measured around $z\approx0.17$ clusters using background quasars with impact parameters within $1.5R_{200}$. The \lya\ $W_{r,500}$ measured for galaxy clusters is significantly lower compared to the $W_{r,500}$ we obtained for the highest mass bin of our sample.

The dotted red line in the plot (right y-axis) shows the star-formation efficiency (SFE) as a function of halo mass. The SFE is defined as ${\rm SFE} \equiv  M_*/M_{\rm baryon}=M_*/(M_{\rm halo}*f_{b})$, where $f_b~ (\equiv \frac{\Omega_{\rm b}}{\Omega_{\rm M}})$ is the mean cosmic baryon fraction. As mentioned earlier, we used the stellar mass -- halo mass (SMHM) relation from \citet[]{Moster_2013} to obtain the SFE curve. 
Both quantities show a peak at some halo mass ($\sim 10^{11}$~\Msun\ for the \lya\ bearing gas and $\sim 10^{12}$~\Msun\ for the SFE) and decline at both the low- and high-mass ends. Such a similarity may suggests that the cool neutral gas content of the CGM has direct consequences for the efficiency of star formation inside galaxies. However, we note that the comparison of a relative quantity such as the SFE with an absolute quantity like the $W_r$ should be done with caution.

 Although the connection of the \lya-bearing gas with the SFE is not straightforward, one particularly important aspect of the right-hand panel of Fig.~\ref{fig:m-b stack_cont} is the increase of $W_r$ with decrease in $M_{\rm halo}$ from $10^{14}$ to $10^{11}~M_{\odot}$. Although the SFE peaks at around $\approx10^{12}~M_{\odot}$ and then decreases with decreasing halo mass, this is not reflected in the \lya\ bearing gas. The lower SFE of galaxies with $M_{\rm halo} \approx10^{11}~M_{\odot}$ compared to $\approx10^{12}~M_{\odot}$ halo mass galaxies, despite being more gas-rich, may be indicative of a larger gas depletion time for these galaxies. Conversely, despite a decreasing \lya\ $W_r$, the increased SFE of galaxies with $M_{\rm halo}\approx10^{12}~M_{\odot}$  compared to $\approx10^{11}~M_{\odot}$ halo mass may suggest that the circumgalactic gas is depleted before the star formation inside galaxies. The recent simulation by \citet[]{Appleby_22} showed that the column density distribution function (CDDF) of \HI\ absorbers around green valley galaxies resembles that around quenched galaxies more closely than that of the star-forming galaxies. They interpreted this as a sign of CGM depletion before the quenching of the host galaxy \citep[see also,][]{Davies_20, Oppenheimer_20}. This is in agreement with our observational findings.

Finally, the suppression of cool, neutral gas around high-mass galaxies can also be a consequence of higher virial temperatures leading to a higher degree of ionization of the CGM. However, the virial temperature corresponding to the intermediate-mass bin is already above the temperature required to collisionally ionize the neutral hydrogen. Simulations predict a transition from almost no virialized gas to substantial virialized gas at $M_{\rm halo}\sim 10^{12}~\rm M_{\odot}$- which is the  transition between cold to hot mode accretion \citep[see e.g.,][]{keres} . The observed suppression of \lya\ absorption around massive galaxies can be a possible indication of this transition. Using a sample of Milky Way-type galaxies in the IllustrisTNG simulation, \citet[]{Ramesh_2022} reported that the CGM of high mass galaxies exhibits less \HI\ gas compared to low-mass galaxies. They attributed this effect to the kinetic mode feedback by SMBH residing in more massive galaxies which can be responsible for sweeping up of the cool gas in the CGM. This is  consistent with our observations.

In passing, we note that our results contrast with \citet{Bordoloi_2018}, who reported a positive trend between \lya\ $W_r$ and $M_*$ using a sample of 85 galaxies with $b<160$ kpc and $8\leq$ \logm\ $\leq$ 11.6. This apparent disagreement may be due to their small sample size, and the fact that they did not take into account the upper limits in their analysis.

\subsection{The role of SFR}

Using a subsample of galaxies with SFR estimates, we studied the impact of SFR and sSFR on the distribution of cool, neutral gas in and around galaxies. The left panel of Fig.~\ref{fig:sfr-b stack} shows a clear indication of enhanced \lya\ absorption within the virial radius for galaxies with high SFR. However, outside the virial radius, no dependence on the SFR is seen. Similarly, in the right panel, a strong SFR dependence is seen only within 100~pkpc. Although we did not explicitly control the redshifts in this analysis, the median redshifts of the compared bins are very similar. The measurements obtained for only the MUSEQuBES galaxies (shown with  green and magenta points) are also consistent with the complete sample.

No statistically significant trend is reported between \lya\ $W_r$ and SFR for the galaxies in the COS-Halos survey \citep[][]{Thom_2012,Tumlinson_2013}. In fact, \citet[]{Thom_2012} argued that the CGM of passive galaxies is equally rich in \HI. However, \citet[][COS-GASS survey]{Borthakur_2015}, found a marginal correlation of individual Ly$\alpha$ absorption strength with SFR using 45 low-redshift galaxies, with the majority of the sightlines passing within the virial radius of the galaxies. Combining the observations of the COS-HALOS survey along with COS-GASS, \citet[]{Borthakur_2016} found a stronger correlation between SFR and impact parameter-corrected \lya\ equivalent width, defined as the ratio of observed \lya\ equivalent width and that predicted by the best-fit model for the $W_r$-profile for the entire sample. Their combined sample is also limited to sightlines within $\approx 1.4~R_{\rm vir}$. This is consistent with the results of our analysis using spectral stacking. A similar trend is also seen for $z\approx3.3$ \lya\ emitters \citep[LAEs; see][]{Muzahid_2021}. However, the LAE sightlines probe the CGM at $b >R_{\rm vir}$, where we did not see any trend with SFR.

The trend between \lya\ absorption around galaxies and SFR can arise from the two following scenarios. First, star-formation driven outflows can entrain and deposit cool gas in the CGM. We recall that the strong SFR dependence is only seen for impact parameters  $\lesssim100$~pkpc. The gas traced by \lya\ at large galactocentric distances does not show any dependence on SFR. This suggests that in this scenario the outflows are only effective in determining the gas distribution out to $\approx100$~pkpc. Second, the availability of more cool gas in the CGM can lead to higher SFRs in galaxies. Although it is not straightforward to relate the gas in the CGM to star-formation activity inside galaxies (since the gas in the CGM has to go through different physical processes via which it can cool and become molecular before forming stars), the second scenario is predicted by simulations \citep[see e.g., ][]{Davies_19, Davies_20}.

 Finally, we note that the median stellar mass of the high-SFR bin is higher than for the low-SFR bin. This is not surprising since the majority of the galaxies in our sample follow the main sequence relation (see Fig.~\ref{fig:prop}). However, the observed trend with SFR is unlikely due to the difference in stellar masses for the two SFR bins, since a similar trend is also seen for sSFR, and again only within the virial radius (see the left panel of Fig.~\ref{fig:ssfr-b stack}). The fact that we see enhanced \lya\ absorption inside the virial radius for star-forming galaxies compared to quenched galaxies (sSFR$<10^{-11}~\rm yr^{-1}$) is consistent with the findings of \citet[]{Johnson_15}, who found an enhanced covering fraction inside the virial radius of late-type galaxies compared to early-type counterparts.

\subsection{Implications of the optical depth maps}

We produced 2D optical depth (OD) maps for the complete sample as well as for two sub-samples with high-mass and low-mass galaxies. In the ideal case of an isotropic distribution of gas around galaxies, circularly symmetric OD maps should be observed. The departure from circular symmetry in the maps indicates the presence of redshift space distortions. This is clear from the enhanced optical depth along the LOS direction at small impact parameters (Fig.~\ref{fig:all stack} and Fig.~\ref{fig:OD map massbin}). However, at larger impact parameters, this effect is less pronounced. Such a distortion along the line of sight direction is also observed by \citet[]{Tejos_14}. They found that the extent of the distortion was consistent with the redshift uncertainty of their galaxy sample. We observed the excess OD up to $\gtrsim 1$~pMpc LOS Hubble distance for the lowest impact parameter bin. Considering a redshift uncertainty of $\approx 50$~\kms\ for our galaxy sample, the length scale of the distortion would be $\sim0.7$~pMpc at the median redshift of $z=0.1$. Hence, the observed elongation of the OD map along the LOS direction unlikely to be entirely due to redshift uncertainty of our galaxy sample. \citet[]{Turner_14} reported a similar redshift space distortion for a sample of LBGs at $z\approx2.3$. Comparing with the {\sc eagle} simulation, \citet[]{Turner_17} argued that infalling gas is responsible for the redshift space distortion. A similar comparison of our observed optical depth maps with simulations could shed light on the gas kinematics around low-redshift galaxies.

The strong excess \lya\ absorption within $\approx100$~pkpc transverse and $\approx1$~pMpc LOS Hubble distance around high-mass galaxies is consistent with our analysis of the $W_r$-profile (right panel of Fig.~\ref{fig:m-b stack}). This is a possible indication of sightlines passing through denser, cool gas clouds originating in high-mass galaxies from galactic processes (e.g., extended galactic disk, clouds from galactic fountains). An increasing strength of \lya\ absorption with stellar mass (and hence halo mass) in the vicinity of galaxies has been predicted by simulations \citep[see e.g.,][]{Rakic_2013,Turner_17}.

The OD maps are qualitatively in agreement with the maps produced by \citet[]{Chen_20} for high-redshift ($z\approx2$) galaxies. However, they find that the total optical depth in the redshifted region of the optical depth map is larger by $>50\%$ than that of the blueshifted side (for $70~<b (\rm kpc) <150$). Producing the OD map without folding the negative and positive LOS velocities we confirmed that this asymmetry is not pronounced in our case. The total optical depth in the redshifted region is only $\approx7$\% higher compared to the blueshifted region. The ``least implausible'' explanation for this asymmetry between redshifted and blueshifted flux according to \citet[]{Chen_20} was that the Ly$\alpha$ emission from galaxy halos contaminates the absorption features. This can be non-negligible if a foreground galaxy is being probed by a background galaxy spectra, as was the case in their study. However, our galaxy sample is probed by much brighter background quasars, so this possible source of contamination can be neglected for all practical purposes. The lack of any  significant asymmetry in the optical depth maps found in this work may be a direct consequence of the aforementioned reason.

 \subsection{Possible caveats}

In this study, we combined galaxy samples from six different surveys from the literature (non-IFS) along with our MUSEQuBES galaxies (IFS) leading to 5054 QSO-galaxy pairs. While such a large number of quasar-galaxy pairs is critical to probe any difference in the \lya\ $W_r$-profile as a function of galaxy properties, it inevitably introduces heterogeneity in the galaxy sample. The different archival surveys have different depths and spectroscopic completeness, and probe a wide range of redshifts ($z\approx 0.01-0.48$ corresponding to 5~Gyr of cosmic time). 
Further, surveys such as COS-Halos and \citet[]{Liang_14} were designed to study the CGM of `isolated' galaxies in contrast with the Keeney+18 survey which includes a substantial fraction of group galaxies in it. Note that because of the lack of SFR measurements, the Keeney+18 sample does not contribute to our analysis of the SFR and sSFR dependence. Hence for the SFR and sSFR dependence, the contributing galaxies are mostly `isolated', and environmental effects are not expected to be important.

We took two measures to minimize the effects of the heterogeneous galaxy sample: (1) Most of our analyses are based on the \lya\ $W_r$-profile as a function of impact parameter and normalized impact parameter ($b/R_{\rm vir}$). This naturally takes care of the different distances from the galaxy centre when the dependence on a given galaxy parameter is investigated. (2) Whenever required, we controlled the redshift of the galaxies in a given $b$ and $b/R_{\rm vir}$ bin (see e.g., Fig.~\ref{fig:m-b stack_cont}).

Finally, another form of bias can be introduced as a consequence of correlated galaxy properties (e.g., between $z$ and $M_*$ and between $M_*$ and SFR). As pointed out earlier, the merging of different samples helps us to mitigate such intertwined correlations (see Fig.~\ref{fig:prop}). We have not explored the effect of galaxy orientation with respect to the quasar sightlines in this work, as it is beyond the scope of this paper.

\section{Summary}
\label{sec:summ}

In this study, we used 4595 galaxies with median $z$ of 0.1 (68\% range of 0.07 -- 0.19) from the MUSEQuBES and archival CGM surveys probed by 184 background quasars to construct \HI\ \lya\ $W_r$-profiles by means of spectral stacking. We report our findings based on median stacks in this work, but we verified that the mean stack produces consistent conclusions. The galaxies span a wide range of stellar mass ($10^{9.1}-10^{10.6}~\rm M_{\odot}$), star formation rate ($0.01-1~\rm M_{\odot}yr^{-1}$), and impact parameter ($0.5-2.5$ pMpc). The impact of different galaxy properties on the \lya\ $W_r$-profile is investigated. Our key findings are: 
 
\begin{itemize}

    \item We find excess \lya\ absorption around low-$z$ galaxies out to a projected distance of 2~pMpc or equivalently $\approx 15 R_{\rm vir}$ and up to a LOS velocity of $\approx \pm 600$ \kms\ in 2D optical depth map and $W_r$-profile. The \lya\ $W_r$-profile is well described by a single power-law with a slope of $\approx -0.80$. (Figs.~ \ref{fig:all stack} \& \ref{fig:b stack}).

    \vskip0.2cm 
    \item The power-law index of the $W_r$-profile is correlated with stellar mass, with a steeper slope for higher mass galaxies (Figs.~ \ref{fig:m-b stack} \& \ref{fig:m-b stack_cont}). The $W_r$-profiles of different mass bins converge at large $b$ but diverge at large $b/R_{\rm vir}$.

    \vskip0.2cm 
    \item Using an $F$--test we found that a log-linear (or Gaussian) + power-law model to the $W_{r,500}$-profile fits better than a single component power-law. The log-linear (or Gaussian) component is only prominent at $b \lesssim R_{\rm vir}$ (Fig.~\ref{fig:lit fit}). We interpret the small-scale and power-law components as representing the 1- and 2-halo terms, respectively.


    \vskip0.2cm 
    \item The \lya\ equivalent width within $0.25<b/R_{\rm vir}<2$ of galaxies peaks at $M_{\rm halo}\sim10^{11}~{\rm M_{\odot}}(M_*\sim10^{9}~\rm M_{\odot})$).   (Fig.~\ref{fig:m-b stack_cont}).

    \vskip0.2cm 
    \item Based on a subsample of 442 galaxy-quasar pairs with SFR measurements, we find \lya\ absorption to be strongly correlated with SFR and sSFR but only within the virial radius (Figs.~\ref{fig:sfr-b stack} \& \ref{fig:ssfr-b stack}).

    \vskip0.2cm 
    \item  The median stacked \lya\ absorption spectrum for the full sample can be modelled by invoking the galaxy-absorber 2-point correlation function of the following form: $(\frac{r}{r_0})^{\gamma}$. We obtained a length scale of $r_0=7.6\pm0.4$~pMpc and a power-law index of $\gamma=-1.57\pm0.05$ for the median stacked \lya\ absorption profile (Fig.~\ref{fig:width_sim}).

\end{itemize} 

Owing to the heterogeneity of the galaxy data used in this study we could not explore the environmental dependence of the cool, neutral gas surrounding galaxies. Using the galaxy sample of Keeney+18 and six of the MUSEQuBES fields with Magellan/MOS follow-up data with a spectroscopic completeness similar to Keeney+18, we will investigate the environmental dependence on the cool, neutral, circumgalactic gas in the future.

\section*{Acknowledgements}

SC gratefully acknowledges support from the European Research Council (ERC) under the European Union’s Horizon 2020 research and innovation programme grant agreement No 864361. LW acknowledges funding by the European Research Council through ERC-AdG SPECMAP-CGM, GA 101020943. SC, SD, and SM acknowledge support from the Indo-Italian Executive Programme of Scientific and Technological
Cooperation 2022–-2024 (TPN: 63673). SD acknowledges Prof. R. Srianand, Dr. Aseem Paranjape, and Pushpak Pandey for insightful discussions.

\section*{Data Availability}

The data underlying this article are available in the ESO (http://archive.eso.org/cms.html) and $HST$ (https://hla.stsci.edu/) public archives.


\bibliographystyle{mnras}
\bibliography{all_ref} 



\appendix

\section{Brief summary of the samples from the literature used in this study}  
 
We carried out literature survey to increase the number of galaxies for stacking. A brief summary of the CGM surveys from which the galaxies are drawn is given below:

\subsection{ {\texorpdfstring{\citet{Liang_14}}{}} } 
\label{sec:A1}

The galaxy sample used in \citet{Liang_14} is prepared by cross-correlating public galaxy and QSO survey data. QSO information was obtained from the HST archive, either COS or STIS. Spectroscopically identified galaxies were searched for in the following public surveys: Nearby Galaxy Catalog (NGC), SDSS, 2MASS, 2dFGRS. They used an impact parameter cut-off of 500 kpc, and a LOS separation of at least 1000 \kms\ from the QSO redshift to avoid QSO proximity effects. We retrieve information about all 195 galaxies used in this work. The galaxies have a median redshift of 0.041 and a wide range of stellar mass ranging from $\approx 10^5~\rm M_{\odot}$ to $\approx 10^{11}~\rm M_{\odot}$. The stellar masses are obtained from NASA-Sloan catalog and the scaling relation between stellar mass and rest-frame absolute $r-$band magnitude. A subsample of galaxies have measured star formation rates using rest-frame UV absolute magnitudes. Out of the 96 QSOs used to probe these galaxies, 13 were observed using STIS. We obtain information about the other 83 QSOs which were observed with COS G130M/G160M gratings from their Table 2.

\subsection{COS-Halos survey}

The COS-Halos survey \citep{Tum_2011,Tumlinson_2013,Werk_2013} was designed to study gaseous halos around 44 low-redshift ($z=0.15-0.35$)  galaxies with stellar masses of $\log_{10} ({\rm M_*}/\rm M_{\odot}) \approx 9.5-11.5$ using 39 UV-bright background quasars with impact parameters $\lesssim150$~kpc ($b/R_{\rm vir} < 0.8$).   

The spectroscopic redshifts for these galaxies are obtained using the LRIS spectrograph at Keck and the MagE spectrograph at Magellan. The stellar masses are estimated from five-band SDSS photometry using a template fitting approach implemented in the {\it kcorrrect} code \citep[][]{Blanton_2017} . The SFRs of these galaxies are obtained from detected nebular emission lines, or limited by their absence. The detected SFR of the galaxy sample ranges from 0.5 $\rm M_{\odot}~yr^{-1}$ to 19 $\rm M_{\odot}~yr^{-1}$ (median 3.23 $\rm M_{\odot}~yr^{-1}$).

\subsection{COS-Dwarf survey}
The COS-Dwarf survey by \citet{Bordoloi_2014} primarily focused on extending the COS-Halos survey to a lower stellar mass limit to understand the CGM around dwarf galaxies. The COS-Dwarf survey was optimized to obtain galaxies with stellar mass less than $10^{10}~\rm M_{\odot}$ at redshift $\leq$ 0.1. This redshift range allowed selecting galaxies based on SDSS spectroscopic catalog. The stellar mass and SFR for the 43 selected galaxies were measured using SDSS photometry. The galaxies used in this work lie within 150 kpc transverse distance from UV bright background QSOs. We retrieve information about all the 43 galaxies and quasars used in this work.

\subsection{COS-GASS survey}
The COS-Gass survey \citep{Borthakur_2015} focused on the connection between the CGM of low-redshift galaxies with the atomic gas content in the ISM probed by the \HI~21cm line. The 45 galaxies used in this study were obtained from GALEX Arecibo SDSS Survey (GASS) with redshift ranging from 0.02 to 0.05, impact parameter $<250$ kpc ($b/R_{\rm vir}=0.2-1.5$) and stellar mass ranging from $10^{10.1}~\rm M_{\odot}-10^{11.1}~\rm M_{\odot}$. Background QSOs probing these galaxies were observed using COS with a limiting flux of FUV$_{mag}\leq$19.0, which yielded 45 usable background QSO. The galaxy redshifts were obtained from SDSS spectroscopic measurements, stellar mass was obtained from the GASS survey \citep{Cat_10}. SFR measurements were obtained from both GALEX FUV and NUV and SDSS photometry. Both star-forming and passive galaxies are present in this sample with the SFR ranging from $10^{-2}~\rm M_{\odot}~yr^{-1}$ to $10^{0.8}~\rm M_{\odot}~yr^{-1}$. We have retrieved information about all the 45 galaxies and QSOs from this work. 

\subsection{{\texorpdfstring{\citet{Johnson_15}}{}}  } 
The galaxy sample used in the study by \citet{Johnson_15} is a combination of 11 galaxies from SDSS, 95 galaxies from their own absorption-blind galaxy survey using the IMACS and LDSS3 spectrographs on the Magellan telescopes. The survey was carried out in four fields, targeting $z<0.4$ galaxies with $r_{AB}<23$ mag and as far as $\Delta \theta=10$ arcmin from the QSO. Information about all of the 106 galaxies are included in our study. We obtain the spectra of the 4 quasars in the 4 fields from the HST-COS archive.  The stellar mass of the galaxy sample is obtained from the g- and r-band absolute magnitude (using multiband photometry and the {\it kcorrect} tool). The galaxy sample spans a stellar mass range of log$_{10}(M_{*}/\rm M_{\odot})$ = 8.4–11.5 with a median of log$_{10}(M_{*}/\rm M_{\odot})$ = 10.3.

\subsection{{\texorpdfstring{\citet{Keeney_2018}}{}} } 
\label{sec:A6}
The galaxy information in \citet{Keeney_2018} is obtained from the COS GTO Galaxy Redshift Survey and the Galaxy Group Survey. The COS GTO Galaxy Redshift Survey obtained redshifts of galaxies with g$<$20 within 38 AGN sight lines, using the HYDRA spectrograph on the WIYN 3.5m telescope and the AA$\Omega$ spectrograph on the 3.9m Anglo-Australian telescope for the multi-object spectroscopy. The Galaxy Group Survey was designed to observe 10 sightlines that probe SDSS selected galaxy groups, with a primary focus on increasing the number of spectroscopically identified group members to $N\geq 20$ per group. Multi-object spectroscopy was performed with WIYN/HYDRA and MMT/Hectospec. We discarded galaxies which have no COS spectrum available or are likely to be a star ($z<0.001$), based on the absorption flag provided in their galaxy catalog. That left us with a total of 8187 galaxies with spectroscopic redshift ranging from 0.001 to 0.909 (median $z=0.149$) around 47 sightlines. The galaxies have a wide range of stellar mass ranging from $\sim 10^5~\rm M_{\odot}$ to $\sim 10^{13}~\rm M_{\odot}$ (median $10^{10.1}~\rm M_{\odot}$). The stellar mass of galaxies in this work is calculated from the galaxy’s rest-frame i-band luminosity. The impact parameters range from $\approx 6$ kpc to $\approx 13$ Mpc with median impact parameter being 2.3 Mpc (median $b/R_{\rm vir} \approx 14  $).

\begin{figure*}
\centering
\begin{subfigure}{0.50\textwidth}

    \centering
    \includegraphics[width=1.0\linewidth]{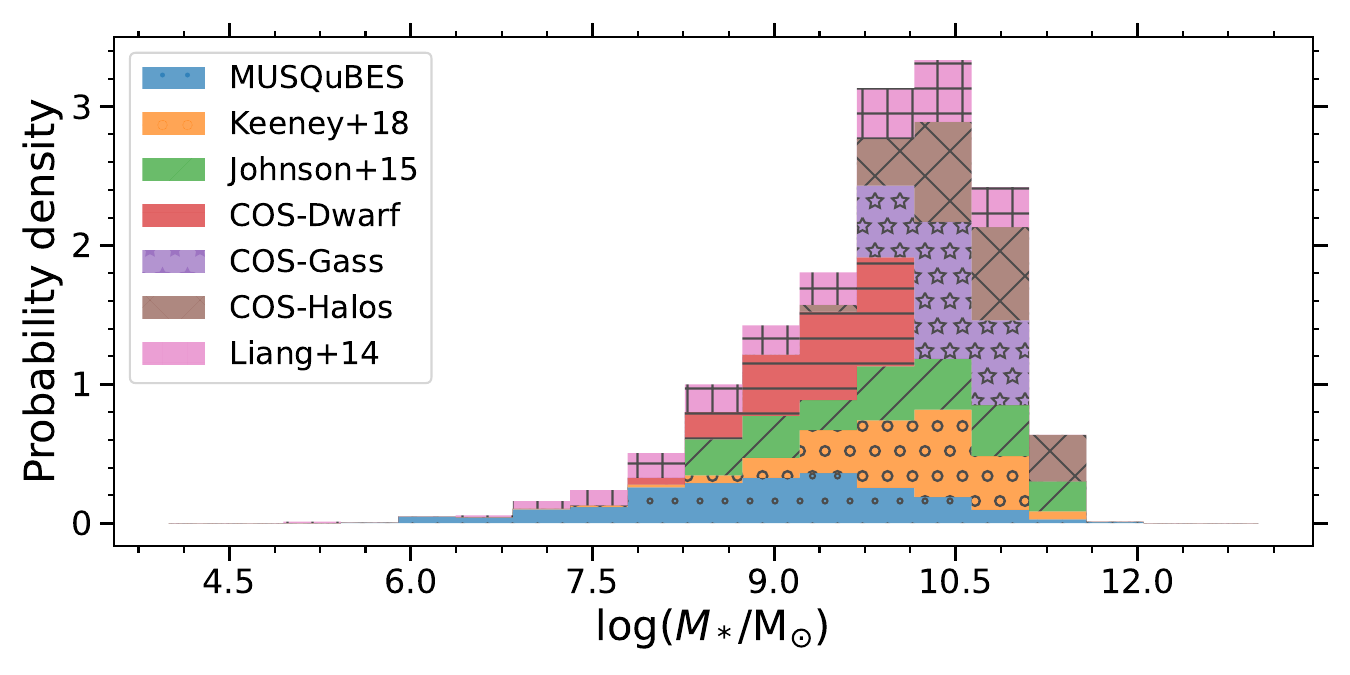}
\end{subfigure}%
\begin{subfigure}{0.50\textwidth}

    \centering
    \includegraphics[width=1.0\linewidth]{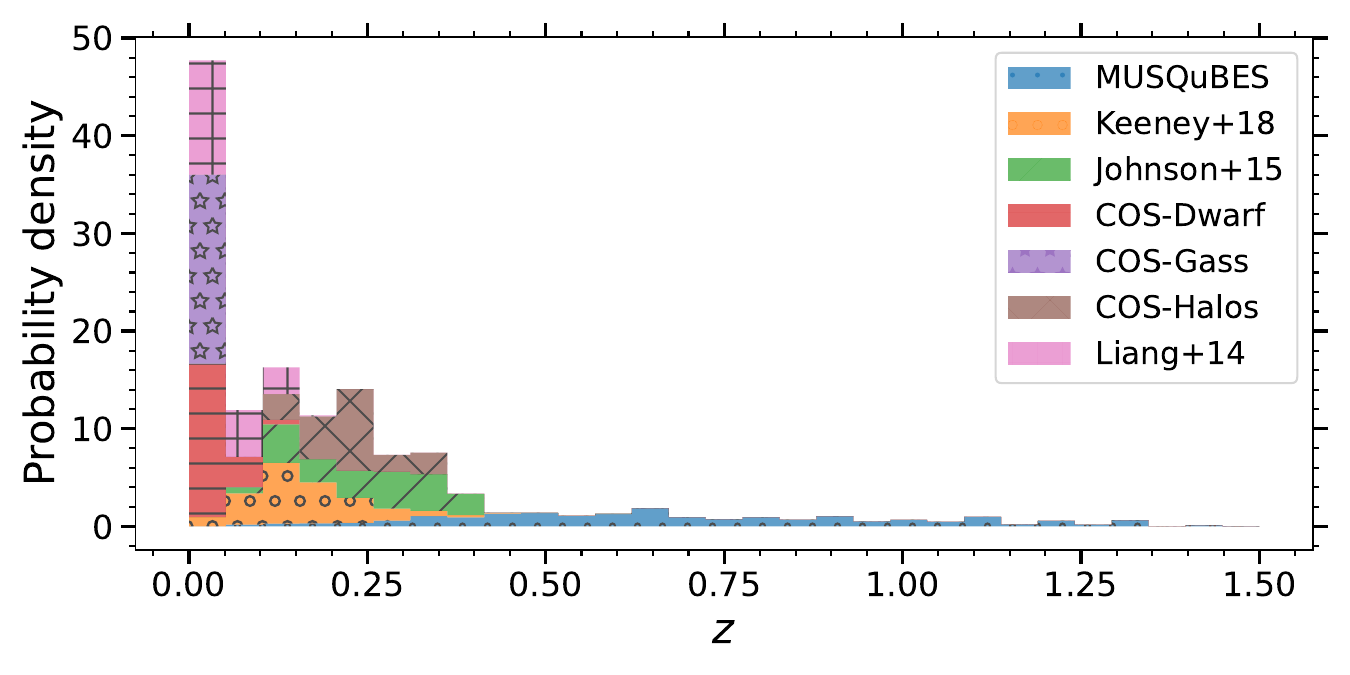}
  
\end{subfigure}
\caption{Probability density distribution of stellar mass (left) and redshift (right) of galaxies in the complete sample, shown with stacked histograms. The stellar mass and redshift distribution of galaxies from different CGM surveys are shown with histograms of different colors and symbols.}
\label{fig:all_mz_dist}
\end{figure*}


\section{Online Only Tables} 
\label{online_only} 

Here we present all the measurements used for the \lya\ $W_r$-profiles presented in this paper (Tables~S1 to S11).  


\bsp	
\label{lastpage}
\end{document}